\documentclass[10pt,twocolumn,twoside]{IEEEtran}
\usepackage{graphicx, multirow}
\usepackage{epsfig}
\usepackage{amssymb}
\usepackage{amsmath}
\usepackage{amstext}

\usepackage{booktabs}
\usepackage{multirow}
\usepackage{array}
\newcolumntype{x}[1]{>{\centering\let\newline\\\arraybackslash\hspace{0pt}}p{#1}}

\usepackage{makecell}
\usepackage[table,xcdraw]{xcolor}

\newcommand{\splitcell}[1]{\begin{tabular}{@{}c@{}}#1\end{tabular}}
\newcommand{\bsplitcell}[1]{$\left[\splitcell{#1}\right]$}

\DeclareMathOperator*{\argmaxA}{arg\,max} 

\usepackage{cite}

\ifCLASSINFOpdf
\else
\fi
\begin{document}
%
\title{VACE-WPE: Virtual Acoustic Channel Expansion Based On Neural Networks for Weighted Prediction Error-Based Speech Dereverberation}
%
%

\author{Joon-Young Yang$^{1}$ and Joon-Hyuk Chang$^2$,~\IEEEmembership{Senior Member,~IEEE}
\thanks{$^1$Hanyang University, Seoul, 04763, Korea (e-mail: dreadbird06@gmail.com).}%
\thanks{$^2$Hanyang University, Seoul, 04763, Korea (e-mail: jchang@hanyang.ac.kr).}%
}

\maketitle

\begin{abstract}
Speech dereverberation is an important issue for many real-world speech processing applications.
Among the techniques developed, the weighted prediction error (WPE) algorithm has been widely adopted and advanced over the last decade, which blindly cancels out the late reverberation component from the reverberant mixture of microphone signals.
In this study, we extend the neural-network-based virtual acoustic channel expansion (VACE) framework for the WPE-based speech dereverberation, a variant of the WPE that we recently proposed to enable the use of dual-channel WPE algorithm in a single-microphone speech dereverberation scenario.
Based on the previous study, some ablation studies are conducted regarding the constituents of the VACE-WPE in an offline processing scenario.
These studies help understand the dynamics of the system, thereby simplifying the architecture and leading to the introduction of new strategies for training the neural network for the VACE.
Experimental results in noisy reverberant environments reveal that VACE-WPE considerably outperforms its single-channel counterpart in terms of objective speech quality and is complementary to the single-channel WPE when employed as the front-end for the far-field automatic speech recognizer.
\end{abstract}


%
\IEEEpeerreviewmaketitle

\section{Introduction}
\label{sec1}
Speech signals traveling in an enclosed space are encountered by walls, floor, ceiling, and other obstacles present in the room, creating multiple reflections of the source image.
Hence, when they are captured by a set of microphones in a distance, the delayed and attenuated replicas of the sound source appear as the so-called reverberation component of the microphone observations.
The reverberation component can be considered a composition of the early reflections and late reverberation \cite{book:room_acoustics}.
In particular, the former is known to change the timbre of the source speech yet helps improve the intelligibility \cite{early_reflections_importance}, whereas the latter degrades the perceptual listening quality as well as deteriorates the performance of speech and speaker recognition applications 
\cite{reverb_challenge_2014,voices_channelge_2019_eval_plan,voices_challenge_2019}.
One of the most popular approaches for speech dereverberation is to exploit the multi-channel linear prediction (MCLP) technique to model the late reverberation component and subsequently cancel it out from the microphone observations.
Specifically, in \cite{Nakatani-WPE-TASLP10}, the delayed linear prediction (LP) model was adopted to estimate the late reverberation, whose model parameters are obtained via iterative maximization of the likelihood function defined under the assumption that the dereverberated speech signal follows a complex normal distribution with time-varying variance.
This method is referred to as the weighted prediction error (WPE) algorithm, and both the time- and short-time Fourier transform (STFT) domain implementations were presented in \cite{Nakatani-WPE-TASLP10};
the latter is usually preferred to the former owing to its computational efficiency.

Several variants of the WPE algorithm or MCLP-based speech dereverberation methods have been proposed for the past decade.
In \cite{Yoshioka-MCLP-TASLP12}, a generalized version of the WPE algorithm \cite{Nakatani-WPE-TASLP10} was derived via the introduction of a new cost function that measures temporal correlation within the sequence of the dereverberated samples.
In \cite{Iwata_log_spectral_priors}, the log-spectral domain priors based on Gaussian mixture models were introduced 
to the procedure for estimating the power spectral density (PSD) of the dereverberated speech signal.
The STFT coefficients of the dereverberated speech were modeled using the Laplacian distribution in \cite{Jukic_laplacian_wpe:is14}, whereas a more general sparse prior, the complex generalized Gaussian (CGG) \cite{complex_generalized_gaussian}, was adopted in \cite{Jukic_sparse_priors}.
More recently, Student's t-distribution was employed as the prior of the desired signal, and the LP filter coefficients were subjected to probabilistic Bayesian sparse modeling with a Gaussian prior \cite{Raj_bayesian_mclp:taslp19}.

Another branch of the WPE variant is to integrate deep neural networks (DNNs) into the WPE-based speech dereverberation framework.
In \cite{neural_wpe:is17}, a DNN was trained to estimate the PSD of the early arriving speech components, which substituted the iterative PSD estimation routine of the conventional WPE algorithm \cite{Nakatani-WPE-TASLP10}.
It was shown in \cite{nwpe_unsup:icassp19} that such a DNN for supporting the WPE algorithm can be trained in an unsupervised manner (i.\,e., without requiring the parallel data for supervision) by performing an end-to-end optimization of the $\ell_{2}$-norm-based cost functions involving the relevant signals.
Moreover, the DNN-supported WPE \cite{neural_wpe:is17} was subjected to an end-to-end joint optimization with a DNN-based acoustic model for robust speech recognition \cite{jt_wpe_online_asr:icassp19}.
Unlike \cite{neural_wpe:is17}, an auto-encoder DNN trained on clean speech was used to constrain the estimated PSD to have characteristics similar to those of the clean speech in a learned feature space \cite{Raj_ae-dnn:eusipco19}.
Meanwhile, a DNN was employed to estimate the shape parameter of the CGG source prior \cite{gwpe_varying_source_priors:waspaa19}, which provides a more flexible form of the WPE algorithm proposed in \cite{Jukic_sparse_priors}. 

A common observation underlying the abovementioned studies 
\cite{Jukic_sparse_priors,Raj_bayesian_mclp:taslp19,neural_wpe:is17,nwpe_unsup:icassp19,jt_wpe_online_asr:icassp19,Raj_ae-dnn:eusipco19} 
is that the multi-channel WPE algorithm is generally superior to its single-channel counterpart.
Inspired by this, we previously proposed the virtual acoustic channel expansion (VACE) technique for the WPE \cite{vace_wpe:is20}, a variant of the WPE designed to utilize the dual-channel WPE algorithm in a single-microphone speech dereverberation scenario.
Specifically, the neural WPE \cite{neural_wpe:is17} is assisted by another neural network that generates the virtual signal from an actual single-channel observation, whereby the pair of actual and virtual signals is directly consumed by the dual-channel neural WPE algorithm.
The neural network for the virtual signal generation, the supposed VACENet, is first pre-trained and then subsequently fine-tuned to produce the dereverberated signal via the actual output channel of the dual-channel neural WPE.

This article is an extension of \cite{vace_wpe:is20}, which aims to provide a more comprehensive understanding of the VACE-WPE based on the empirical evaluation results obtained via sets of experiments, each of which is designed to investigate the dynamics of the VACE-WPE with respect to the various system constituents.
The limitations of the previous study \cite{vace_wpe:is20} are listed below:
\begin{itemize}
\item The VACE-WPE system in \cite{vace_wpe:is20} was designed rather ad hoc, and the dynamics of the system was not sufficiently investigated.
\item Because \cite{vace_wpe:is20} is essentially a feasibility study, the experiments were conducted only in the noiseless reverberant conditions, which is practically unrealistic.
\end{itemize}
Accordingly, the contribution of this article is two-fold:
\begin{itemize}
\item Some ablation studies are conducted with regard to the system components of the VACE-WPE, which helps understand the characteristics of the VACE-WPE and further leads to an overall performance improvement.
\item Experimental results in noisy reverberant environments are provided, 
which demonstrates that the VACE-WPE is significantly superior to the single-channel WPE in achieving better objective speech quality, while both being complementary with each other as the front-end for the reverberant speech recognition task.
\end{itemize}

\section{Overview of the VACE-WPE} \label{sec3}

\subsection{Signal Model} \label{sec3:A}
Suppose that a speech source signal is captured by $D$ microphones in a reverberant enclosure.
In the STFT domain, the observed signal impinging on the $d$-th microphone can be approximated as follows \cite{Nakatani-WPE-TASLP10,Yoshioka-MCLP-TASLP12}:
\begin{align}
{X}_{t,f,d} &= \sum_{\tau=0}^{l-1}{h_{\tau,f,d}^{*}}{S_{t-\tau,f}} + V_{t,f,d}\,,
\end{align}
where $S_{t,f}$ and $V_{t,f,d}$ denote the STFT-domain representations of the source speech and noise observed at the $d$-th microphone, respectively; 
the superscript $*$ denotes the complex conjugate operation, and $h_{t,f,d}$ represents the room impulse response (RIR) from the source to the $d$-th microphone, whose duration is $l$.
Further decomposing the speech term into the early arriving component (i.\,e., the direct path plus the early reflections) and late reverberation \cite{Nakatani-WPE-TASLP10} provides
\begin{align}
{X}_{t,f,d} &= \sum_{\tau=0}^{\Delta-1}{h_{\tau,f,d}^{*}}{S_{t-\tau,f}} + \sum_{\tau=\Delta}^{l-1}{h_{\tau,f,d}^{*}}{S_{t-\tau,f}} + V_{t,f,d} \\
			&= {X}_{t,f,d}^{\textrm{(early)}} + {X}_{t,f,d}^{\textrm{(late)}} + V_{t,f,d}\,, \label{eqn:signal_model}
\end{align}
where $\Delta$ denotes the STFT-domain time index and determines the duration of the RIR that contributes to the early arriving speech component.
Herein, the early arriving speech is assumed to be obtained upon convolution between the source speech and the RIR truncated up to 50 ms after the main peak.
%
Accordingly, 
with the 64 ms Hann window and a hop size of 16 ms employed for the STFT analysis, 
$\Delta$ is fixed to 3 (16$\times$3$\,\approx\,$50).

\subsection{Review of the WPE Algorithm} \label{sec3:B}

\subsubsection{Iterative WPE}
Under the noiseless assumption that ${V}_{t,f,d}=0$, $\forall{d}$, the late reverberation component, ${X}_{t,f,d}^{\textrm{(late)}}$, in Eq.\,(\ref{eqn:signal_model}) can be approximated by the delayed LP technique as follows \cite{Nakatani-WPE-TASLP10}:
\begin{align}
\hat{X}_{t,f,d}^{\textrm{(late)}} &= \sum_{\tau=\Delta}^{\Delta+K-1}{\textbf{g}_{\tau,f,d}^{H}}{{\textbf{X}}_{t-\tau,f}} \\
								  &= {\tilde{\textbf{g}}_{f,d}^{H}}{{\tilde{\textbf{X}}}_{t-\Delta,f}}\,,
\end{align}
where ${\textbf{g}_{\tau,f,d}} \in \mathbb{C}^{D}$ represents the $K$-th order time-invariant LP filter coefficients for the output channel index $d$; 
${{\textbf{X}}_{t,f}} \in \mathbb{C}^{D}$ represents the $D$-channel stack of the microphone input signal; 
${\tilde{\textbf{g}}_{f,d}} = {[ {\textbf{g}_{\Delta,f,d}^{T}}, ..., {\textbf{g}_{\Delta+K-1,f,d}^{T}} ]}^{T} \in \mathbb{C}^{DK}$, ${\tilde{\textbf{X}}_{t-\Delta,f}} = {[ {\textbf{X}_{t-\Delta,f}^{T}}, ..., {\textbf{X}_{t-(\Delta+K-1),f}^{T}} ]}^{T} \in \mathbb{C}^{DK}$, and $T$ and $H$ denote the hermitian and transpose operations, respectively.
Under the assumption that ${X}_{t,f,d}^{\textrm{(early)}}$ is sampled from a complex normal distribution with a zero mean and time-varying variance, $\lambda_{t,f,d}$, the objective of the WPE algorithm is to maximize the log-likelihood function \cite{Nakatani-WPE-TASLP10,Yoshioka-MCLP-TASLP12}:
\begin{gather}
{\tilde{\textbf{g}}_{f,d}^{\prime}}, \lambda_{t,f,d}^{\prime} = \argmaxA_{{\tilde{\textbf{g}}_{f,d}}, \lambda_{t,f,d}} L_{f,d}\,, \\
L_{f,d} = \mathcal{N}(\hat{X}_{t,f,d}^{\textrm{(early)}} = {X}_{t,f,d} - {\tilde{\textbf{g}}_{f,d}^{H}}{{\tilde{\textbf{X}}}_{t-\Delta,f}} ; \, 0,\,\lambda_{t,f,d})\,
\end{gather}
for $d \in \{1, 2, ..., D\}$.
As this optimization problem has no analytic solution, ${\tilde{\textbf{g}}_{f,d}}$ and $\lambda_{t,f,d}$ are alternatively updated via the following iterative procedure \cite{Nakatani-WPE-TASLP10,Yoshioka-MCLP-TASLP12}:
\begin{align}
\textrm{Step 1)}
~~~& {\lambda}_{t,f} = \frac{1}{D} \sum_{d}^{} \left( \frac{1}{2\delta+1} \sum_{\tau=-\delta}^{\delta} {\lvert {{Z}_{t+\tau,f,d}} \rvert}^2 \right), \label{eqn:wpe_step1}
\end{align}
\begin{alignat}{3}
\textrm{Step 2)}
~~~~~~& \textbf{R}_{f} = \sum_{t}^{} \frac{{\tilde{\textbf{X}}}_{t-\Delta,f} {\tilde{\textbf{X}}_{t-\Delta,f}^{H}}} {{\lambda}_{t,f}} && \in \mathbb{C}^{DK \times DK}, \label{eqn:wpe_step2}\\
~~~~~~& \textbf{P}_{f} = \sum_{t}^{} \frac{{\tilde{\textbf{X}}}_{t-\Delta,f} {{\textbf{X}}_{t,f}^{H}}} {{\lambda}_{t,f}} && \in \mathbb{C}^{DK \times D}, \\
~~~~~~& \textbf{G}_{f} = {\textbf{R}_{f}^{-1}} {\textbf{P}_{f}} && \in \mathbb{C}^{DK \times D}, \label{eqn:wpe_step2-3} \\
\textrm{Step 3)}
~~~~~~& {{\textbf{Z}}_{t,f}} = {\textbf{X}}_{t,f} - \textbf{G}_{f}^{H} {\tilde{\textbf{X}}}_{t-\Delta,f}, && \label{eqn:wpe_step3}
\end{alignat}
where Eq.\,(\ref{eqn:wpe_step1}) is obtained by further assuming that ${\lambda}_{t,f,1} = {\lambda}_{t,f,2} = ... = {\lambda}_{t,f,D}$, and $\delta$ is the term introduced to consider the temporal context between the neighboring frames.
$\textbf{G}_{f}$ is a matrix whose $d$-th column is ${\tilde{\textbf{g}}_{f,d}}$, and ${{\textbf{Z}}_{t,f}} =\hat{\textbf{X}}_{t,f,d}^{\textrm{(early)}}$ is the $D$-channel stack of the dereverberated output signal.
In the first iteration, ${{\textbf{Z}}_{t,f}}$ is initialized to ${{\textbf{X}}_{t,f}}$.
It was revealed in \cite{Yoshioka-MCLP-TASLP12} that the WPE algorithm described in Eqs.\,(\ref{eqn:wpe_step1})\,--\,(\ref{eqn:wpe_step3}) can be derived as a special case of the generalized WPE, without enforcing the noiseless assumption.

\subsubsection{Neural WPE}
Neural WPE \cite{neural_wpe:is17} exploits a neural network to estimate the PSD of the 
dereverberated output signal, ${\lvert {{Z}_{t,f,d}} \rvert}^2$, as follows:
\begin{align}
\ln{{\lvert {\hat{Z}_{t,f,d}} \rvert}^2} = \mathcal{F} \left( {\ln{{\lvert {{X}_{d}} \rvert}^2}}; \Theta_\textrm{LPS} \right), \label{eqn:lpsnet}
\end{align}
where $\mathcal{F}(\,\cdot\,; \Theta_\textrm{LPS})$ denotes the neural network parameterized by $\Theta_\textrm{LPS}$, to estimate the log-scale power spectra (LPS) of the dereverberated signal in a channel-independent manner;
the time-frequency (T-F) indices were dropped in $X_{d}$, as neural networks often consume multiple T-F units within a context as the input.
Accordingly, Eq.\,(\ref{eqn:wpe_step1}) can be rewritten as follows:
\begin{align}
{\lambda}_{t,f} = \frac{1}{D} \sum_{d}^{} {\lvert {\hat{Z}_{t,f,d}} \rvert}^2. \label{eqn:nwpe_step1}
\end{align}

For the rest of this paper, we will denote the neural network for the PSD estimation, $\mathcal{F}(\,\cdot\,; \Theta_\textrm{LPS})$, as the LPSNet \cite{vace_wpe:is20}, as it operates in the LPS domain of the relevant signals.

\begin{figure}[t]
\begin{center}
	\centerline{\includegraphics[width=3.27in]{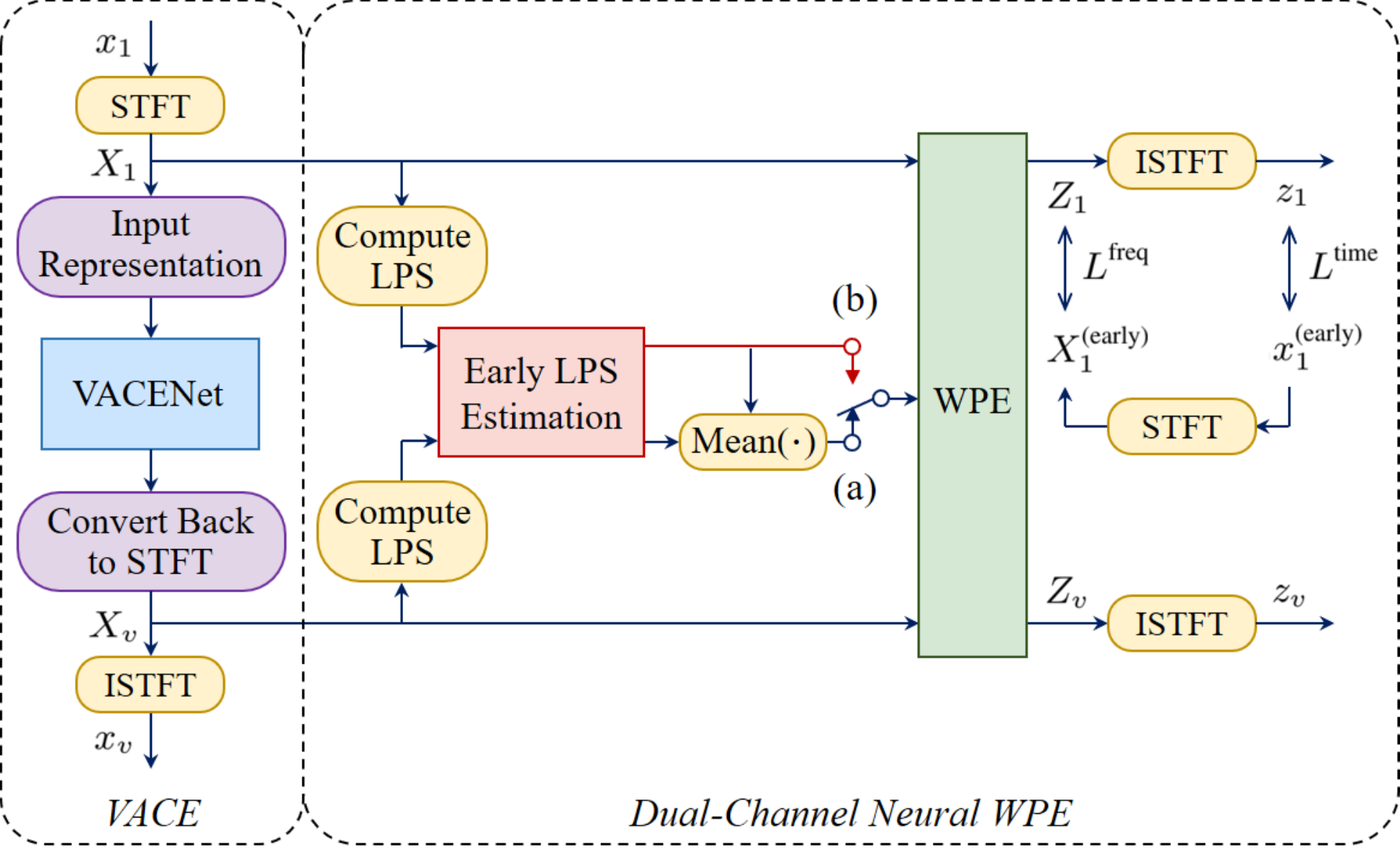}}
	\caption{
		Block diagram of the VACE-WPE systems: (a) VACE-WPE \cite{vace_wpe:is20} and (b) simplified VACE-WPE.
		The subscripts $1$ and $v$ denote the actual and virtual channel signals, respectively.}
	\label{fig:vace_wpe}
\end{center}
\end{figure}

\subsection{VACE-WPE System Description} \label{sec3:C}

\subsubsection{Overview} \label{sec3:C:1}
The entire VACE-WPE system \cite{vace_wpe:is20} consists of two separate modules: the VACE module, which is responsible for the generation of the virtual signal, and the dual-channel neural WPE, which operates in the exact same manner as described in Eqs.\,(\ref{eqn:wpe_step2})\,--\,(\ref{eqn:nwpe_step1}) for $D=2$.
To build the complete VACE-WPE system, the LPSNet is trained to estimate the LPS of the early arriving speech given the reverberant observation, and the VACENet is pre-trained under a certain predefined criterion.
These two steps are independent of each other, and thus, can be performed in parallel.
Subsequently, the VACE-WPE system is constructed as depicted in Fig.\,\ref{fig:vace_wpe}, and the VACENet is fine-tuned to produce the dereverberated signal at the output channel corresponding to the actual microphone.
During the fine-tuning, the LP order is fixed to $K=K_\textrm{trn}$, and the parameters of the LPSNet are frozen.

\begin{figure}[t]
	\begin{center}
		\centerline{\includegraphics[width=\linewidth]{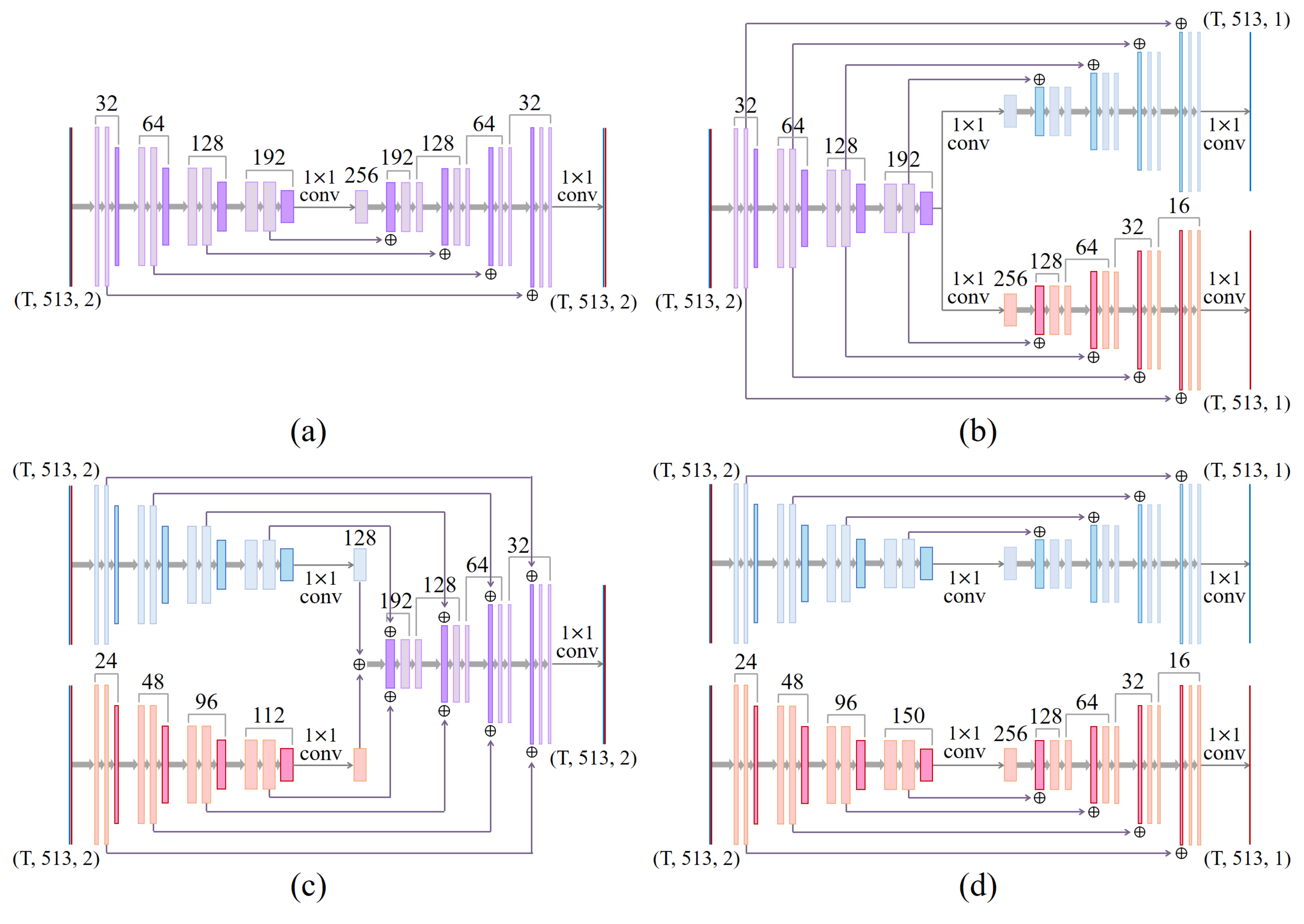}}
		\caption{
			Four different VACENet architectures for modeling the RI components: 
			(a) VACENet-a, (b) VACENet-b, (c) VACENet-c, and (d) VACENet-d. 
			The input and output feature maps are represented in \textit{(time}, \textit{frequency}, \textit{channel)} format, and the numbers above the rectangles denote the number of channels.}
		\label{fig:vacenet_architectures}
	\end{center}
\end{figure}

\subsubsection{Architecture of the VACENet} \label{sec3:C:2}
Similar to our previous study \cite{vace_wpe:is20}, we used the U-Net \cite{unet2015} as the backbone architecture of the VACENet, 
whose input and output representations are the real and imaginary (RI) components of the STFT coefficients of the actual and virtual signals, respectively.
Unlike \cite{vace_wpe:is20}, four different architectures of the VACENet are considered in this study, 
each of which differs in whether to use a shared or a separate stream for the convolutional encoder and decoder.
Fig.\,\ref{fig:vacenet_architectures} shows the detailed illustration of the four distinctive VACENet architectures, denoted as VACENet-\{a,\,b,\,c,\,d\}.
First, all the models consume both of the RI components as the input for the encoder stream, whether it is separated or not, which is
intended to fully exploit the information residing in the pair of the RI components.
Second, the VACENet-\{a,\,c\} use a shared decoder stream to model the RI components of the virtual signal, whereas the VACENet-\{b,\,d\} split the decoder stream into two to separately model each attribute of the RI components.
As shown in Fig.\,\ref{fig:vacenet_architectures}, the difference between the VACENet-b and VACENet-d lies in whether the separated decoder streams share the bottleneck feature or not, as well as the encoder feature maps for the skip connections.
Meanwhile, VACENet-c can be considered a more flexible version of the VACENet-a, as it splits the encoder stream into two separate streams, and thus, doubles the number of skip connections originating from the encoder module. 

In each subfigure in Fig.\,\ref{fig:vacenet_architectures}, the rectangles denote the feature maps, 
whose height and width represent their relative size and depth, respectively, 
and the numbers above the rectangles are the channel sizes of the feature maps.
Each of the wide arrows denotes a 2D convolution (Conv2D) with a kernel size of 3, and $\boldsymbol{\oplus}$ denotes the concatenation of the feature maps along the channel axis.
Every downsampling or upsampling operation is either performed by a $3\times3$ Conv2D or a transposed Conv2D with a stride size of 2, and $1\times1$ convolutions are used in the bottleneck and the last layers of the network.
A gated linear unit \cite{glu:icml17} was used instead of a simple convolution followed by an activation function, except for the layers for downsampling and upsampling.
Lastly, to make fair comparisons between the different model structures, we designed each model to have a similar number of parameters in total, as shown in Table \ref{tab:vacenet_params}.

A similar investigation regarding the model architecture was conducted in \cite{tan_crn:icassp19} for the speech enhancement task, where the structure analogous to that depicted in Fig.\,\ref{fig:vacenet_architectures}-(b) was shown to be effective.
In contrast, it was mentioned in \cite{multi_metrics_se} that separately handling each RI component is beneficial.
Because the existing task, and hence the role of the VACENet, is fundamentally different from that of the neural networks adopted for speech enhancement \cite{tan_crn:icassp19,multi_metrics_se}, we argue that it is worthwhile to examine which architecture is more appropriate for the VACE task.

\renewcommand{\arraystretch}{1.2}
\begin{table}[]
\centering
\caption{Model size of the different VACENet architectures}
\label{tab:vacenet_params}
\begin{tabular}{|c|x{1.0cm}|x{1.22cm}|x{1.0cm}||x{0.95cm}|}
	\hline
	\multirow{2}{*}{\begin{tabular}[c]{@{}c@{}}Model Config.\end{tabular}} & \multicolumn{4}{c|}{\#parameters (in millions)}  \\ \cline{2-5} 
	& Encoder & Bottleneck & Decoder & Total \\ \hline\hline
	VACENet-a  & 2.21M  & 0.05M  & 3.38M  & 5.64M \\ \hline
	VACENet-b  & 2.21M  & 0.10M  & 3.14M  & 5.45M \\ \hline
	VACENet-c  & 1.93M  & 0.03M  & 3.69M  & 5.65M \\ \hline
	VACENet-d  & 2.60M  & 0.08M  & 2.85M  & 5.53M \\ \hline
\end{tabular}
\end{table}

\subsubsection{Loss Function} \label{sec3:C:3}
Two types of loss functions, namely the frequency-domain loss and time-domain loss, are defined to train the VACENet \cite{vace_wpe:is20}:
\begin{align}
\begin{split}
{L}^\textrm{freq}(A, B) &= \alpha\cdot [\textrm{MSE}(A^{r}, B^{r}) + \textrm{MSE}(A^{i}, B^{i})] \\
&\quad + \beta\cdot \textrm{MSE}(\textrm{ln}|A|, \textrm{ln}|B|), \label{eqn:L1_freq}
\end{split} \\
{L}^\textrm{time}(a, b) &= \textrm{MAE}(a, b), \\
{L}^\textrm{}(A, B) &= {L}^\textrm{freq}(A, B) + \gamma\cdot {L}^\textrm{time}(a, b), \label{eqn:L1_tot}
\end{align}
where $A$ and $B$ are the STFT coefficients,
$\textrm{ln}|A|$ and $\textrm{ln}|B|$ are the log-scale magnitudes; 
$a$ and $b$ are the time-domain signals obtained by taking the inverse STFT of $A$ and $B$, respectively; 
the superscripts $r$ and $i$ denote the RI components, respectively; 
$\alpha$, $\beta$, and $\gamma$ are scaling factors to weigh the losses defined in different domains of the signal representations, and MSE$(\cdot,\cdot)$ and MAE$(\cdot,\cdot)$ compute the mean squared and absolute error between the inputs, respectively.

It is worth noting that $\alpha$ and $\beta$ should be determined such that the values of $\alpha\cdot [\textrm{MSE}(A^{r}, B^{r}) + \textrm{MSE}(A^{i}, B^{i})]$ and $\beta\cdot \textrm{MSE}(\textrm{ln}|A|, \textrm{ln}|B|)$ are similar.
When the former is considerably larger than the latter, severe checkerboard artifacts \cite{checkerboard_effect_cnn} were revealed in the output signal of the network.
For the opposite condition, it was not able to obtain fine-grained representations of the RI components of the output signal.
$\gamma$ was also set to make $\gamma\cdot {L}^\textrm{time}(a, b)$ to have values similar to or slightly smaller than those of the aforementioned two terms.

\subsubsection{Pre-training of the VACENet} \label{sec3:C:4}
In this study, we consider two different pre-training strategies to initialize the VACENet.
Suppose that the time-domain representations of the actual and virtual signals are denoted by $x_1$ and $x_v$, respectively, and their STFT-domain counterparts $X_1$ and $X_v$, respectively.
Then, the forward pass of VACENet can be expressed as follows:
\begin{align}
X_v = \mathcal{G} \left( X_1; \Theta_\textrm{VACE} \right),
\end{align}
where  $\mathcal{G}(\,\cdot\,; \Theta_\textrm{VACE})$ denotes the VACENet parameterized by $\Theta_\textrm{VACE}$.
First, considering the observed signal as the input, the VACENet can be pre-trained to reconstruct the input signal itself \cite{vace_wpe:is20} by minimizing the loss function $L(X_v, X_1)$.
Alternatively, we propose to pre-train the VACENet 
to estimate the late reverberation component of the input signal, denoted by $X_1^{\textrm{(late)}}$, by minimizing $L(X_v, X_1^{\textrm{(late)}})$.

The rationale behind the invention of these pre-training strategies is rather simple and intuitive. 
Under the assumption that the actual dual-channel speech recordings may not deviate significantly from each other, we employed the first method in \cite{vace_wpe:is20}, while expecting the virtual signal to resemble the observed signal.
However, the generated virtual signal was shown to have characteristics different from the observed signal \cite{vace_wpe:is20}, and the shape and scale of the waveform resembled those of the late reverberation component of the observed signal, as shown in Fig.\,\ref{fig:plot_AB_sf_late} in Section \ref{sec5:C}.
Accordingly, we suggest initializing VACENet to produce the late reverberation component of the observed signal.
For the rest of this paper, we denote the two pre-training strategies described above as \textit{PT-self} and \textit{PT-late}.

\subsubsection{Fine-tuning of the VACENet} \label{sec3:C:5}
As mentioned earlier, VACENet is fine-tuned within the VACE-WPE architecture depicted in Fig.\,\ref{fig:vace_wpe}.
The loss function is set to $L(Z_1, X_1^\textrm{(early)})$, where $X_1^\textrm{(early)}$ denotes the early arriving speech component of the observed signal, $X_1$, and $Z_1$ is the output of the WPE algorithm on the actual channel side \cite{vace_wpe:is20};
the virtual channel output, $Z_v$, is neglected.

\subsubsection{Simplification of the PSD Estimation Routine} \label{sec3:C:6}
In addition to the architecture of the original VACE-WPE system \cite{vace_wpe:is20} depicted in Fig.\,\ref{fig:vace_wpe}-(a), 
we propose the simplified VACE-WPE, depicted in Fig.\,\ref{fig:vace_wpe}-(b), by removing the contribution of the virtual signal to the PSD estimation routine expressed in Eq.\,(\ref{eqn:nwpe_step1}).
Accordingly, Eq.\,(\ref{eqn:nwpe_step1}) can be rewritten as follows:
\begin{align}
{\lambda}_{t,f} = {\lvert {\hat{Z}_{t,f,1}} \rvert}^2.
\label{eqn:nwpe_step1_simp}
\end{align}

One of the motivations behind this modification is to take away some burden from the roles of the VACENet by reducing the dependency of the model to the entire system.
In other words, if we consider the WPE-based dereverberation as a two-stage process of early arriving speech PSD estimation (Eq.\,(\ref{eqn:lpsnet})) followed by decorrelation (Eqs.\,(\ref{eqn:wpe_step2})\,--\,(\ref{eqn:wpe_step3})), the VACENet in Fig.\,\ref{fig:vace_wpe}-(a) is expected to generate the virtual signal whose role is to contribute to both the stages.
In contrast, as the contribution of the virtual signal to the first stage is removed in Fig.\,\ref{fig:vace_wpe}-(b), the VACENet would concentrate more on the second stage.
Further details regarding the simplified VACE-WPE system are provided in Section \ref{sec5:B} with the experimental results.

\section{Experimental Setup} \label{sec4}
%


\subsection{On-the-fly Data Generator} \label{sec4:A}
To present as many random samples as possible to the neural networks during the training, an on-the-fly data generator was used.
Given the sets of clean speech utterances, RIRs, and noises, the data generator first randomly selects a speech utterance, an RIR, and a noise sample from each set, respectively.
Then, the speech utterance is randomly cropped, and subsequently convolved with the full-length RIR as well as the truncated RIR to create the reverberated speech and early arriving speech, respectively.
The noise sample is either cropped or duplicated to match the duration of the speech excerpt and added to both the reverberated and early arriving speech; 
the signal-to-noise ratio (SNR) is randomly chosen within the predefined range of integers.

\renewcommand{\arraystretch}{1.25}
\begin{table}[]
	\centering
  \caption{Parameters for RIR simulation \cite{kaldi_rirs} based on image method \cite{image_method_rir}}
	\label{tab:simulated_rirs}
	\begin{tabular}{|c|c|c|c|}
		\hline
		\multicolumn{2}{|c|}{Parameter}                                              & Medium           & Large            \\ \hline\hline
		\multirow{2}{*}{\begin{tabular}[c]{@{}c@{}}Room size\end{tabular}} & lower bound & {[}10$\times$10$\times$2{] $\textrm{m}^{3}$}    & {[}30$\times$30$\times$2{] $\textrm{m}^{3}$}    \\ \cline{2-4} 
		& upper bound & {[}30$\times$30$\times$5{] $\textrm{m}^{3}$}    & {[}50$\times$50$\times$5{] $\textrm{m}^{3}$}    \\ \hline
		\multicolumn{2}{|c|}{Duration}                                               & 1.0 s            & 2.0 s            \\ \hline
		\multicolumn{2}{|c|}{Reflection order}                                       & \multicolumn{2}{c|}{10}             \\ \hline
		\multicolumn{2}{|c|}{Absorption coefficient}                                 & \multicolumn{2}{c|}{{[}0.2, 0.8{]}} \\ \hline
		\multicolumn{2}{|c|}{Source-Receiver distance}                               & \multicolumn{2}{c|}{{[}1.0, 5.0{] \textrm{m}}} \\ \hline
	\end{tabular}
\end{table}

\subsection{Training Datasets} \label{sec4:B}
\subsubsection{TrainSimuClean} \label{sec4:B:1}
The clean speech utterances were taken from the ``training" portion of the TIMIT \cite{timit} dataset, which comprises phonetically balanced English speech 
sampled at 16 kHz.
After excluding the common-transcript utterances and filtering out those with durations of less than 2 s, we obtained 3,337 utterances from 462 speakers;
the average duration of the training utterances was 3.21 s.
The simulated RIRs in \cite{kaldi_rirs} were used for the training, which is freely available\footnote{https://www.openslr.org/28/} and widely used in Kaldi's speech and speaker recognition recipes for data augmentation purposes \cite{kaldi_toolkit}.
A total of 16,200 medium room and 5,400 large-room RIRs were randomly selected to construct a simulated RIR dataset for the training, where we excluded the small room RIRs to check whether the trained neural WPE variants can generalize well to the small room conditions at the evaluation time.
The parameters of the RIR simulation \cite{image_method_rir} are presented in Table \ref{tab:simulated_rirs}, and further details can be found in \cite{kaldi_rirs}.
No additive noise samples were used in this dataset.
\subsubsection{TrainSimuNoisy} \label{sec4:B:2}
The modified LibriSpeech-80h dataset was used as the clean speech corpus, which is a subset of the LibriSpeech \cite{librispeech} corpus and provided as part of the VOiCES Challenge 2019 dataset \cite{voices_channelge_2019_eval_plan,voices_challenge_2019}.
It consists of read English speech sampled at 16 kHz, whose transcripts are derived from public domain audiobooks.
As most of the speech samples contain considerable amounts of epenthetic silence regions as well as those at the beginning and end of the utterance, we employed an energy-based voice activity detector implemented in Kaldi \cite{kaldi_toolkit} to trim the silence regions.
The utterances whose duration was less than 2.8 s were filtered out after the silence removal.
Consequently, we obtained 16,341 utterances from 194 speakers, with an average speech duration of 12.26 s.
The simulated RIR dataset described in Section \ref{sec4:B:1} was reused.
As for the noise dataset, we used 58,772 audio samples in the DNS Challenge 2020 dataset \cite{dns_challenge_2020}, which contains audio clips selected from Google Audioset\footnote{https://research.google.com/audioset} and Freesound\footnote{https://freesound.org}.
The dataset comprises 150 unique audio classes, including animal sounds, vehicular sounds, indoor and outdoor environment sounds originating from various things and daily supplies, music of different genres, and musical instruments.
%

Instead of directly feeding the raw clean speech samples to the neural network models during the training, we set a limit on the dynamic range of the speech waveform amplitudes as described in the following.
Suppose that $\mathbf{x}$ is a vector of the time-domain speech waveform amplitudes normalized to have values between -1 and 1.
Then, the waveform amplitudes after applying a simple dynamic range control (DRC) scheme can be obtained as follows:
\begin{align}
\mathbf{x}_\textrm{drc} = \mathbf{x} \cdot \frac{2}{\bar{a}_\textrm{max}-\bar{a}_\textrm{min}} \cdot r,
\end{align}
where $\bar{a}_\textrm{max}$ and $\bar{a}_\textrm{min}$ are the average of the $n$ largest and $n$ smallest waveform amplitudes, respectively, and $r$ is a constant for the DRC;
$n=100$ and $r=0.25$ were used in this study.

\renewcommand{\arraystretch}{1.25}
\begin{table}[]
\centering
\caption{Specifications of the real RIRs taken from the REVERB Challenge 2014 dataset \cite{reverb_challenge_2014}}
\label{tab:reverb14_rirs}
\begin{tabular}{|c|c|c|c|}
	\hline
	Condition            & Duration            & $T_{60}$                & Recording distance \\ \hline\hline
	\textit{Small-near}  & \multirow{6}{*}{1s} & \multirow{2}{*}{0.25 s} & 0.5 m              \\ \cline{1-1} \cline{4-4} 
	\textit{Small-far}   &                     &                         & 2 m                \\ \cline{1-1} \cline{3-4} 
	\textit{Medium-near} &                     & \multirow{2}{*}{0.5 s}  & 0.5 m              \\ \cline{1-1} \cline{4-4} 
	\textit{Medium-far}  &                     &                         & 2 m                \\ \cline{1-1} \cline{3-4} 
	\textit{Large-near}  &                     & \multirow{2}{*}{0.75 s} & 0.5 m              \\ \cline{1-1} \cline{4-4} 
	\textit{Large-far}   &                     &                         & 2 m                \\ \hline
\end{tabular}
\end{table}

\subsection{Test Datasets} \label{sec4:C}
\subsubsection{TestRealClean}
The ``core test" set of the TIMIT \cite{timit} dataset was used as the clean speech corpus, where no speakers and transcripts overlap with those of the \textit{TrainSimuClean} dataset described in Section \ref{sec4:B:1};
the average speech duration is 3.04 s.
The entire set of utterances was randomly convolved with the real RIRs taken from the REVERB Challenge 2014 \cite{reverb_challenge_2014} dataset to create six unique test sets, each of which differs in the room size as well as the recording distance for the RIR measurement.
Among the eight microphone channels \cite{reverb_challenge_2014}, only the first and fifth channels were used to create the dual-channel test sets; these two channels were located on the opposite side of each other at a distance of 20 cm.
The specifications of the real RIRs are presented in Table \ref{tab:reverb14_rirs}.
Similar to \textit{TrainSimuClean}, \textit{TestRealClean} contains no additive noise.
\subsubsection{TestRealNoisy}
To create the \textit{TestRealNoisy} dataset, the stationary air conditioner noise residing in each room \cite{reverb_challenge_2014} as well as the nonstationary babble and factory noise from the NOISEX-92 \cite{noisex-92} dataset and the music samples from the MUSAN \cite{musan_corpus} dataset were added to the \textit{TestRealClean} dataset.
To simulate test environments with various SNR levels, the noise samples were added to the reverberated speech with the SNRs randomly chosen between 5 dB and 15 dB.

\renewcommand{\arraystretch}{1.25}
\begin{table}[]
	\caption{LPSNet architecture adopted and modified from \cite{pirhosseinloo2019}}
	\label{tab:lpsnet_architecture}
	\centering
	\begin{tabular}{|c|c|c|c|}
		\hline
		Layer                & Kernel     & Stride & \#channels  \\ \hline\hline
		Conv2D + Bias + GLU  & 5$\times$5 & 1$\times$1 & 32      \\ \hline
		Conv2D + Bias        & 5$\times$5 & 1$\times$2 & 32      \\ \hline
		Conv2D + Bias + GLU  & 5$\times$5 & 1$\times$1 & 48      \\ \hline
		Conv2D + Bias        & 5$\times$5 & 1$\times$2 & 48      \\ \hline
		Reshape              & -          & -          & 6,192   \\ \hline
		\rowcolor[HTML]{EFEFEF}
		Conv1D + BN + ELU    & 3          & 1          & 256     \\ \hline
		\cellcolor[HTML]{EFEFEF}\bsplitcell{DilatedConv1DBlock \\ Conv1D + BN + ELU} $\times$~4 & \cellcolor[HTML]{EFEFEF}3 & \cellcolor[HTML]{EFEFEF}1 & \cellcolor[HTML]{EFEFEF}256  \\ \hline
		Shortcut Sum + ReLU & -           & -          & 256     \\ \hline
		Conv1D              & 1           & 1          & 513     \\ \hline
	\end{tabular}
\end{table}

\renewcommand{\arraystretch}{1.15}
\begin{table}[]
	\caption{Structure of the DilatedConv1DBlock}
	\label{tab:dconvblock}
	\centering
	\begin{tabular}{|c|c|c|c|c|}
		\hline
		Layer                                                                     & Kernel & Stride & Dilation & \#channels \\ \hline\hline
		\begin{tabular}[c]{@{}c@{}}Conv1D\_$k$ + ELU\\ (for $k=1, ..., 6$)\end{tabular} & 3      & 1      & $\textrm{2}^{k}$       & 16         \\ \hline
		Conv1D + ELU                                                              & 3      & 1      & 1        & 256        \\ \hline
		Sigmoid                                                                   & -      & -      & -        & -          \\ \hline
	\end{tabular}
\end{table}

\renewcommand{\arraystretch}{1.25}
\begin{table}[]
	\caption{Hyperparameters for training the LPSNet models}
	\label{tab:lpsnet_training}
	\centering
	\begin{tabular}{|c|x{2.0cm}|x{2.0cm}|}
		\hline
		Dataset                   & \textit{TrainSimuClean} & \textit{TrainSimuNoisy}  \\ \hline\hline
		Mini-batch size, duration & 4, \,{[}2.0, 2.8{]}\,s    & 6, \,{[}2.4, 2.8{]}\,s     \\ \hline
		SNR range                 & -                       & {[}3, 20{]} dB           \\ \hline
		\#iters\,/\,epoch         & 6,000                   & 12,000                   \\ \hline
	\end{tabular}
\end{table}

\subsection{LPSNet Specifications} \label{sec4:D}
We adopted the dilated convolutional network proposed in \cite{pirhosseinloo2019} as the LPSNet architecture, but with a few modifications.
Tables \ref{tab:lpsnet_architecture} and \ref{tab:dconvblock} show the detailed architecture of the LPSNet and DilatedConv1DBlock, respectively, where the latter works as a building block for the former.
In Table \ref{tab:lpsnet_architecture}, ``BN" is the batch normalization \cite{batchnorm}, ``ELU" is the exponential linear unit \cite{elu_2015}, and ``Shortcut Sum" takes the summation of the outputs of the layers in the shaded rows.
In Table \ref{tab:dconvblock}, a feature map is first processed by a stack of dilated Conv1D layers and another Conv1D layer, and further compressed to have values between 0 and 1 using the sigmoid function. 
This compressed representation is element-wise multiplied to the feature map fed to the DilatedConv1DBlock, thus working as an analogue to a T-F mask.
Note that the input LPS features were also normalized using a trainable BN \cite{batchnorm}.

The LPSNet was trained for 65 epochs using the Adam optimizer \cite{adam_optimizer}, where the initial learning rate was set to $10^{-4}$ and halved after the 20th, 35th, 45th, and 55th epochs.
Dropout regularization \cite{dropout} was applied with a drop rate of 0.3 for every third mini-batch, and gradient clipping \cite{gradient_clipping} was used to stabilize the training with a global norm threshold of 3.0.
The weights of the LPSNet were also subject to $\ell_{2}$-regularization with a scale of $10^{-5}$.
The specifications regarding the mini-batch composition and the number of iterations defined for a single training epoch are presented in Table \ref{tab:lpsnet_training}.

\subsection{VACENet Specifications} \label{sec4:E}
The architecture of the VACENet is basically the same as that of the U-Net \cite{unet2015}, including the number of downsampling and upsampling operations and positions of the concatenations between the encoder and decoder feature maps.
Similar to the LPSNet, each attribute of the input RI components was normalized using a trainable BN \cite{batchnorm}. 
In addition, the RI components of the output signal were de-normalized using the pre-computed mean and variance statistics.
Other details of the VACENet are described in Section \ref{sec3:C:2} and Fig.\,\ref{fig:vacenet_architectures}.

The training of the VACENet was conducted in a manner similar to that described in Section \ref{sec4:D} for training the LPSNet, employing the same on-the-fly mini-batching scheme presented in Table \ref{tab:lpsnet_training}.
Table \ref{tab:vacenet_hparams} shows the hyperparameters set during the pre-training and fine-tuning of the VACENet models, where the values of $\alpha$, $\beta$, and $\gamma$ were determined by monitoring the first few thousand iterations of the training.
To make fair comparisons across the different VACE-WPE systems, all the VACENet models were trained for 60 epochs, both in the pre-training and fine-tuning stages.
In the pre-training stage, the learning rate was initially set to $10^{-4}$ and annealed by a factor of 0.2 after the 20th and 40th training epochs, whereas in the fine-tuning stage, the initial learning rate was set to $5\cdot10^{-5}$ and annealed in the same manner.

\subsection{Evaluation Metrics} \label{sec4:F}
The dereverberation performance of the WPE algorithms was evaluated in terms of the perceptual evaluation of speech quality (PESQ) \cite{P.862.2}, cepstrum distance (CD), log-likelihood ratio, frequency-weighted segmental SNR (FWSegSNR) \cite{se_eval_metrics}, and non-intrusive normalized signal-to-reverberation modulation energy ratio (SRMR) \cite{non_intrusive_srmr}.
For the metrics computation, the early arriving speech was used as the reference signal, except for the SRMR, which can be calculated from the processed signal itself.

\renewcommand{\arraystretch}{1.2}
\begin{table}[]
	\caption{Hyperparameters for training the VACENet models. $p$ denotes the dropout rate.}
	\label{tab:vacenet_hparams}
	\centering
	\begin{tabular}{|c|c|c|c|c|c|}
		\hline
		Dataset   & \multicolumn{3}{c|}{\textit{TrainSimuClean}} & \multicolumn{2}{c|}{\textit{TrainSimuNoisy}} \\ \hline\hline
		\multirow{2}{*}{Stage} & \multicolumn{2}{c|}{Pre-training} & \multirow{2}{*}{\begin{tabular}[c]{@{}c@{}}Fine-\\[-3pt] tuning\end{tabular}} & Pre-training & \multirow{2}{*}{\begin{tabular}[c]{@{}c@{}}Fine-\\[-3pt] tuning\end{tabular}} \\ \cline{2-3} \cline{5-5}
		& PT-self & PT-late &       & PT-late &      \\ \hline
		$\alpha$  & 10      & 10      & 10    & 2       & 1    \\ \hline
		$\beta$   & 0.3     & 0.1     & 0.1   & 0.05    & 0.1  \\ \hline
		$\gamma$  & 20      & 20      & 20    & 10      & 5    \\ \hline
		$p$       & 0.5     & 0.3     & 0.3   & 0.3     & 0.3  \\ \hline
		\#iters\,/\,epoch & \multicolumn{3}{c|}{6,000} & 12,000 & 9,000 \\ \hline
	\end{tabular}
\end{table}

\section{Experimental Results and Analysis} \label{sec5}

In this section, the experimental results and analysis of the VACE-WPE system are provided.
The ablation studies regarding the constituents of the VACE-WPE are provided from Section \ref{sec5:A} to \ref{sec5:D}; 
these studies are performed under noiseless reverberant conditions;
that is, the LPSNet and VACENet models are trained on \textit{TrainSimuClean} and evaluated on \textit{TestRealClean}.
The rationale behind this design of experiments is that, by excluding any interferences other than reverberation, it would be easier to observe how the different system components of the VACE-WPE influence the operating characteristics of the system as well as the realization of the virtual signal.
The results of noisy reverberant conditions and speech recognition results on real recordings are provided in Section \ref{sec5:E} and Section \ref{sec5:F}, respectively.

The baseline systems under comparison are the single- and dual-channel neural WPE algorithms, where the latter is fed with actual dual-channel speech signals;
for the latter, only the dereverberated signal at the first output channel will be under evaluation.
Although it is not possible to exploit the dual-channel WPE in a single-microphone speech dereverberation scenario, it was included for comparison purposes.
Please note that the results for the iterative WPE \cite{Nakatani-WPE-TASLP10,Yoshioka-MCLP-TASLP12} are not presented, as it requires a cumbersome process of parameter tuning, for example, the context parameter, $\delta$, in Eq.\,(\ref{eqn:wpe_step1}) and the number of iterations, per test condition;
nevertheless, the performance of the iterative WPE was slightly worse than that of the neural WPE, when measured on our test datasets.

\begin{figure}[t]
	\begin{center}
		\centerline{\includegraphics[width=\linewidth]{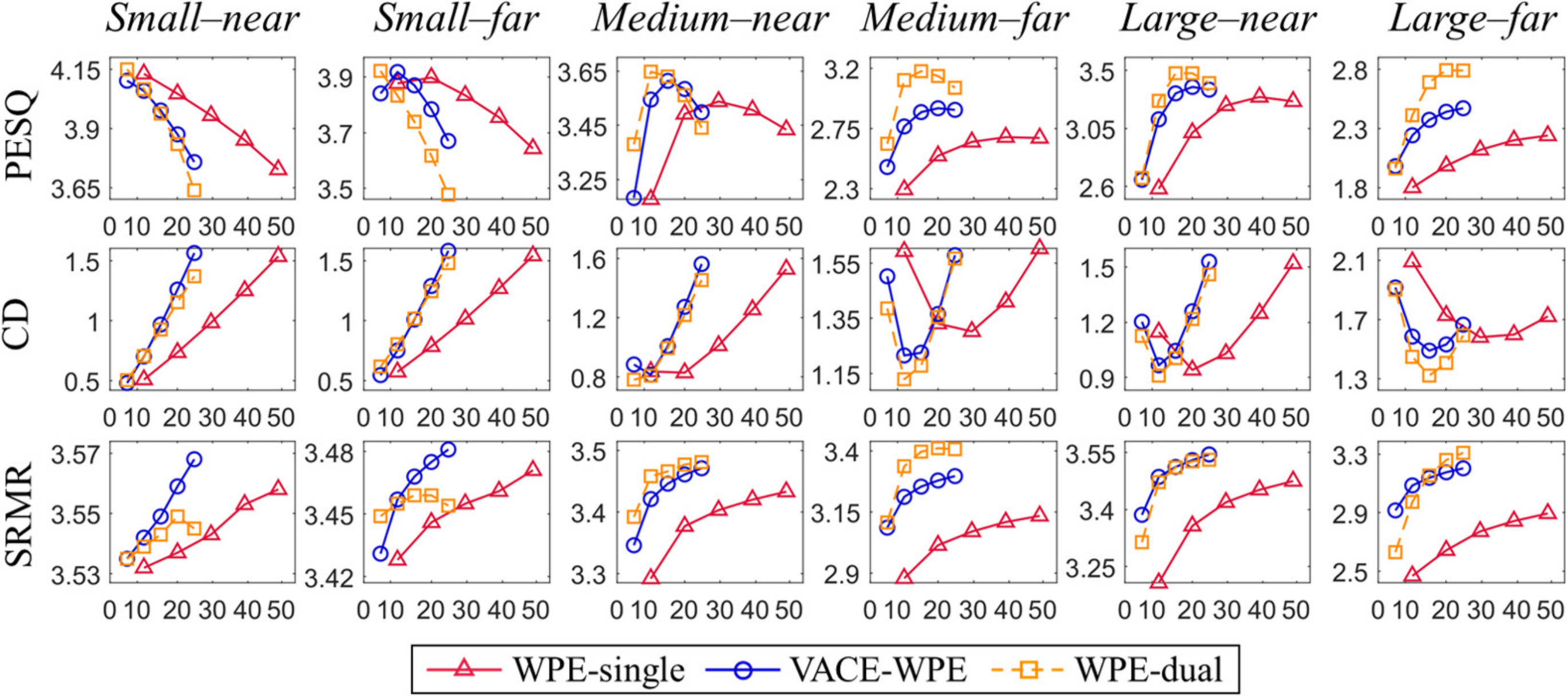}}
		\caption{
			Performance evaluation results of the VACE-WPE and baseline WPE algorithms on \textit{TestSimuClean}.
			The horizontal axis denotes the LP filter order, $K$.
			The VACE-WPE employed the VACENet-b model pre-trained with the PT-self method, and was constructed as depicted in Fig.\,\ref{fig:vace_wpe}-(a).
			$K_\textrm{trn}$ was set to 10 during fine-tuning.
		}
		\label{fig:plot_A}
	\end{center}
\end{figure}

\subsection{Comparison to the Baselines} \label{sec5:A}
\subsubsection{Performance Analysis} \label{sec5:A:1}
Similar to our previous study \cite{vace_wpe:is20}, we first compared the VACE-WPE with the baseline single- and dual-channel WPE algorithms.
To start with the VACE-WPE that has an architecture identical to that described in \cite{vace_wpe:is20}, 
the VACENet-b was pre-trained using the PT-self method and fine-tuned within the VACE-WPE architecture, as depicted in Fig.\,\ref{fig:vace_wpe}-(a), with $K_\textrm{trn}$ set to 10.
Fig.\,\ref{fig:plot_A} demonstrates the evaluation results on \textit{TestSimuClean} in terms of the PESQ, CD, and SRMR metrics.
As shown in the figure, the evaluation for each algorithm was conducted over the fixed sets of LP orders having a constant step size, 
that is, $K\in\{10,\,20,\,30,\,40,\,50\}$ and $K\in\{5,\,10,\,15,\,20,\,25\}$ for the single-channel WPE and dual-channel versions, respectively.
Although these values may not represent the best operating points, it is sufficient to observe the performance variation of each algorithm across the different values of the LP order and to compare the overall performance of the different WPE-based dereverberation methods.

First, in the small room conditions, as the LP order grows, the PESQ score monotonically decreased while the CD increased.
This is because large LP orders lead to overestimation of reverberation, and thus, to speech distortion in a room with a low reverberation time ($T_{60}$).
In contrast, the SRMR slightly increased with $K$, as it only considers the energy ratio in the modulation spectrogram \cite{non_intrusive_srmr}, and thus, cannot accurately reflect the distortions relative to the reference signal.
All three methods revealed the lowest CD at their smallest considered LP orders, exhibiting overall comparable performance.

In the medium room conditions, the performance measured at a far distance was certainly inferior to that measured in the near distance.
Moreover, setting $K$ too small or large led to inaccurate estimation of late reverberation, as demonstrated by both the PESQ and CD metrics.
Unlike the observations in the small room conditions, there are noticeable performance gaps between the single-channel WPE and the others, which are further emphasized in the far distance condition.
Furthermore, there are operating points at which the VACE-WPE outperforms the single-channel WPE in terms of all three metrics, yet is not competitive with the dual-channel WPE.
The results in the large room conditions showed patterns similar to those observed in the medium rooms, but with overall performance degradation, which is attributed to the increased reverberation level.

\begin{figure*}[!t]
	\begin{center}
		\centerline{\includegraphics[width=5.1in]{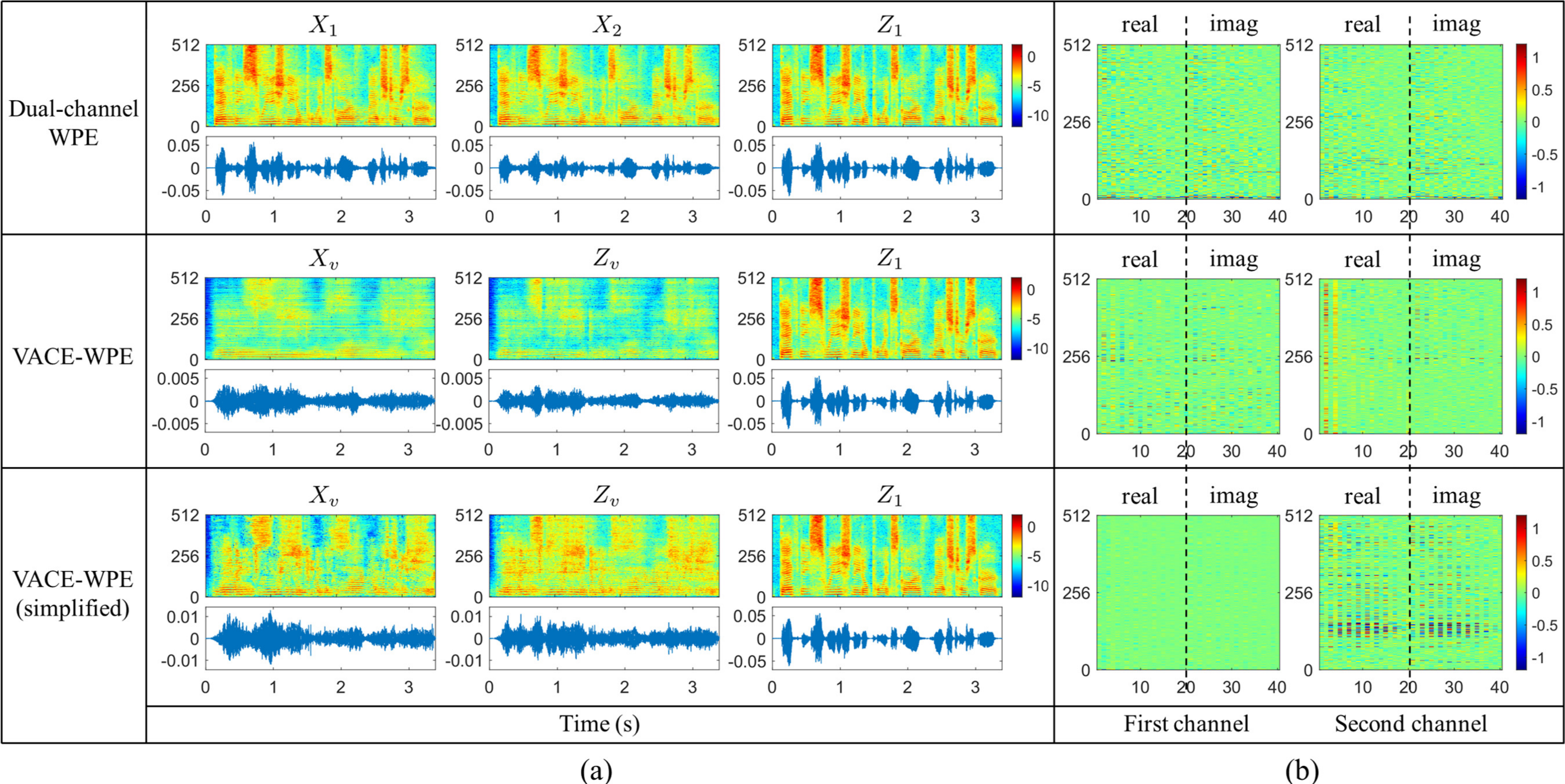}}
		\caption{
			(a) Spectrograms and waveforms of the input and output signals of the different WPE algorithms; 
			$X_1$ and $X_2$ denote the actual first and second channel signals, $X_v$ is the virtual signal; 
			and $Z_1$ and $Z_v$ denote the WPE output signals corresponding to $X_1$ and $X_v$, respectively.
			(b) Visualization of (complex-valued) LP filters ($K=10$) of the WPE algorithms. 
			The label ``First channel" denotes the filter applied to the first channel input signal.
			In each subfigure of the filter, the left and right halves represent the real and imaginary components, respectively.
		}
		\label{fig:plot_AB_sf_sigfilt}
	\end{center}
\end{figure*}

\subsubsection{Visualization of Virtual Signals and LP Filters} \label{sec5:A:2}
As both the dual-channel WPE and VACE-WPE in \cite{vace_wpe:is20} share the same neural WPE back-end, but only differ in the type of the secondary input signal, we compared the input and output signals of the two systems.
Fig.\,\ref{fig:plot_AB_sf_sigfilt} shows the spectrograms and waveforms and the LP filter coefficients obtained from a sample test utterance taken from \textit{TestRealClean} in the \textit{Large-near} condition; 
the filters were calculated with $K=10$.
%
As shown in the first two rows, the generated virtual signal ($X_v$) appears to be considerably different from the pair of actual signals ($X_1$ and $X_2$), yet the dereverberated outputs ($Z_1$'s) look similar.
This implies that, other than the actual observation, an alternative form of the secondary signal that facilitates blind dereverberation via Eqs.\,(\ref{eqn:wpe_step2})\,--\,(\ref{eqn:nwpe_step1}) exists, and a mechanism for generating such a signal can be learned in a data-driven manner using a neural network.
A noticeable feature of the virtual signal is the scale difference, where the amplitudes of the waveform were reduced by an approximate factor of 0.1, as shown in Fig.\,\ref{fig:plot_AB_sf_sigfilt}.
This ``amplitude shrinkage" started to appear in the very early stage of the fine-tuning, even though the VACENet was initialized using the PT-self method to produce the signals whose amplitudes are similar to those of the inputs.
We conjecture that this may be attributed to setting the LP order, $K_\textrm{trn}$, to a constant during the fine-tuning, which forces the VACENet to generate virtual signals that can effectively function as the secondary input for the WPE operating with a fixed LP order, regardless of the degree of reverberation measured in the observed signal.
Nonetheless, it can be seen from the rightmost panel of Fig.\,\ref{fig:plot_A} that the VACE-WPE does not break down when the LP order at the inference time does not match with that employed for the fine-tuning.

The LP filter coefficients of the dual-channel WPE and VACE-WPE, with $K$ set to 10, are demonstrated in the right panel of Fig.\,\ref{fig:plot_AB_sf_sigfilt}.
This clearly verifies that, despite the same operations expressed by Eqs.\,(\ref{eqn:wpe_step2})\,--\,(\ref{eqn:nwpe_step1}), the principles behind the late reverberation estimation are completely different between the two algorithms.
For example, the filters of the dual-channel WPE for both channels seem to focus more on the low-frequency bands, whereas those of the VACE-WPE \cite{vace_wpe:is20} are concentrated on some specific frame delay indices over a wide range of frequency bins and reveal more inter-channel asymmetry.

In terms of perceptual quality, an informal listening test revealed that the virtual signal does not necessarily sound like a completely natural speech, playing machine-like sounds occasionally.
This was attributed to the checkerboard artifacts \cite{checkerboard_effect_cnn}, which inevitably appeared in some utterances.
In addition, the virtual signal sounded more like a delayed and attenuated version of the observed speech, similar to the late reverberation component.
Accordingly, the phonetic sounds or pronunciations of the linguistic contents still remained to some extent, but not as clear as those contained in the original utterance.

\begin{figure}[t]
	\begin{center}
		\centerline{\includegraphics[width=\linewidth]{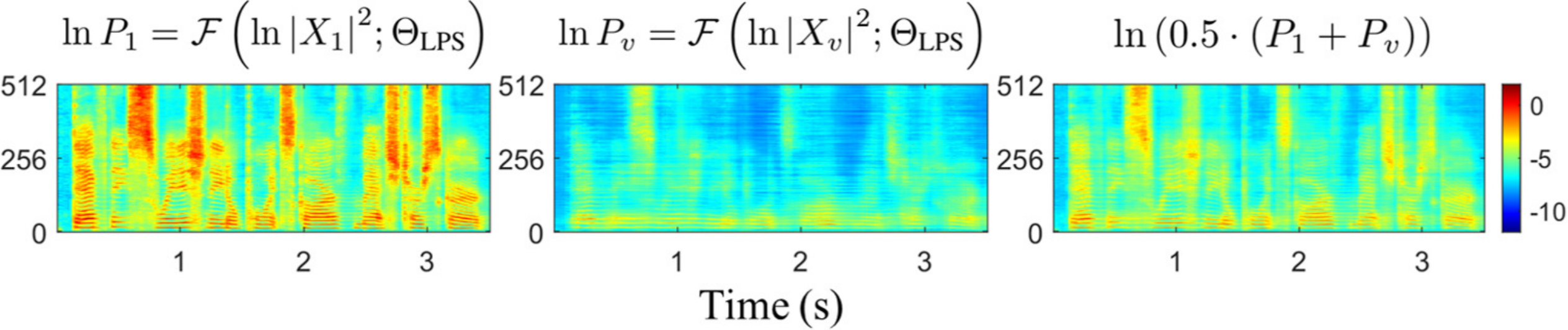}}
		\caption{Spectrograms (in log-magnitudes) obtained from the output of the LPSNet.}
		\label{fig:plot_AB_sf_lpsnet}
	\end{center}
\end{figure}

\begin{figure}[t]
\begin{center}
	\centerline{\includegraphics[width=\linewidth]{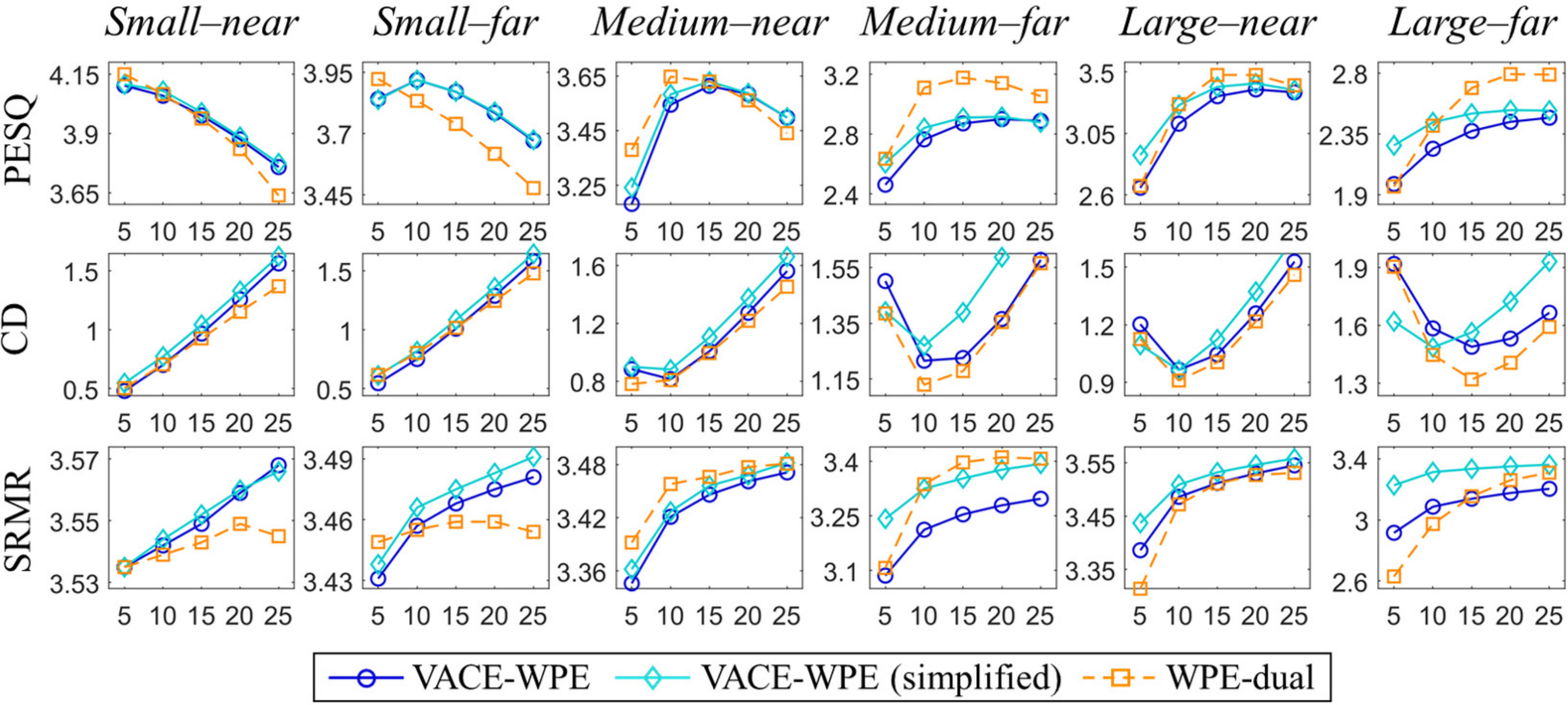}}
	\caption{
		Performance comparison between the VACE-WPE systems before and after the simplification of the PSD estimation routine described in \ref{sec3:C:6}.
		The horizontal axis denotes the LP filter order, $K$.
		Both systems share the same VACENet-b model pre-trained with the PT-self method.
		$K_\textrm{trn}$ was set to 10 during fine-tuning.
	}
	\label{fig:plot_B}
\end{center}
\end{figure}

\subsection{Simplification of the PSD Estimation Routine} \label{sec5:B}
An observation regarding the LPSNet, derived from the ``amplitude shrinkage" of the virtual signal, is shown in Fig.\,\ref{fig:plot_AB_sf_lpsnet}.
In the figure, the first two images are the outputs of the LPSNet, given the actual and virtual signals as the inputs, respectively, and the last image is the average PSD obtained via Eq.\,(\ref{eqn:nwpe_step1}).
As seen in the figure, due to the significant reduction in the amplitudes of the virtual signal, followed by the channel-wise average operation in Eq.\,(\ref{eqn:nwpe_step1}), the average PSD is merely faded out from the power scale of the reverberated or dereverberated speech of the reference (actual) channel.
Based on this observation, we hypothesized that this fadeout would adversely affect the operation of the VACE-WPE, thereby modifying the system. architecture, as depicted in Fig.\,\ref{fig:vace_wpe}-(b).
Section \ref{sec3:C:6} further explains the simplified architecture.

Fig.\,\ref{fig:plot_B} shows the comparisons between the VACE-WPE in \cite{vace_wpe:is20} and the simplified VACE-WPE in terms of the PESQ, CD, and SRMR metrics.
Herein, the simplified VACE-WPE was constructed by fine-tuning the pre-trained VACENet-b, described in Section \ref{sec5:A:1}, within the simplified architecture; 
the same hyperparameters were employed for the fine-tuning.
Note that we omitted the results for the single-channel WPE for visual clarity.
Overall, the simplification boosted both the PESQ and SRMR scores, particularly in the \textit{Medium-far} and \textit{Large-far} conditions by considerable margins, with marginal increments in the CD measures.
In other words, it can be regarded that the simplified VACE-WPE has become better capable of fitting to larger rooms and farther distance conditions, at the expense of slight increase in CD.
The spectrograms and waveforms of the virtual signals related to the simplified VACE-WPE are presented in the last row of Fig.\,\ref{fig:plot_AB_sf_sigfilt}.
Relative to the system without the simplification, the LP filters seem to exploit the virtual signal more aggressively.
Meanwhile, the amplitudes of the virtual signals were amplified by an approximate factor of 2.0.

For the rest of the sections, we use the simplified architecture for all the experiments.

\begin{figure}[t]
	\begin{center}
		\centerline{\includegraphics[width=\linewidth]{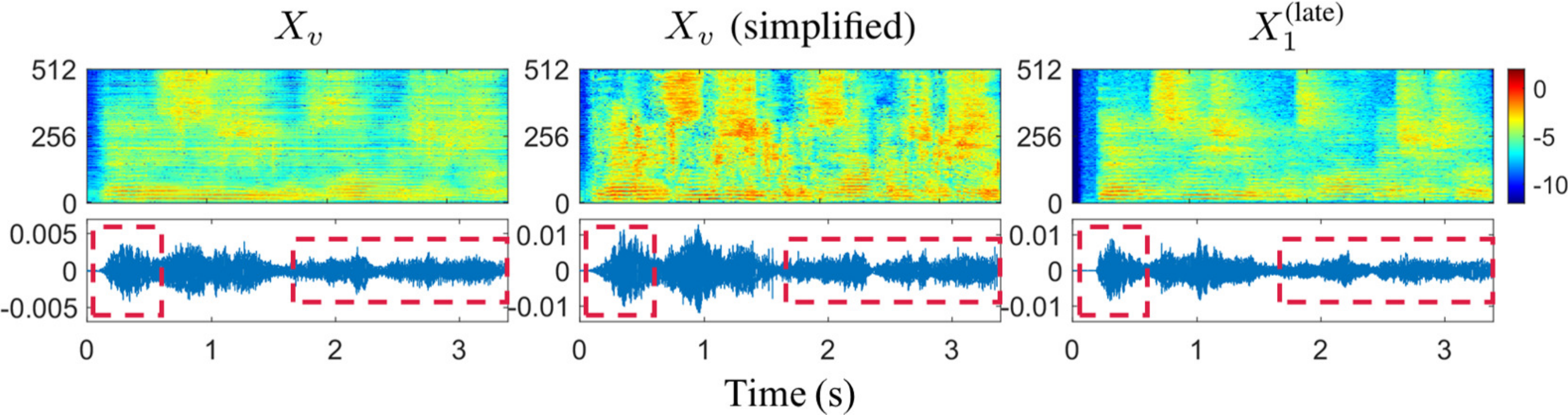}}
		\caption{Spectrograms and waveforms of the virtual signals and the oracle late reverberation signal.}
		\label{fig:plot_AB_sf_late}
	\end{center}
\end{figure}

\begin{figure*}[!t]
	\begin{center}
		\centerline{\includegraphics[width=\linewidth]{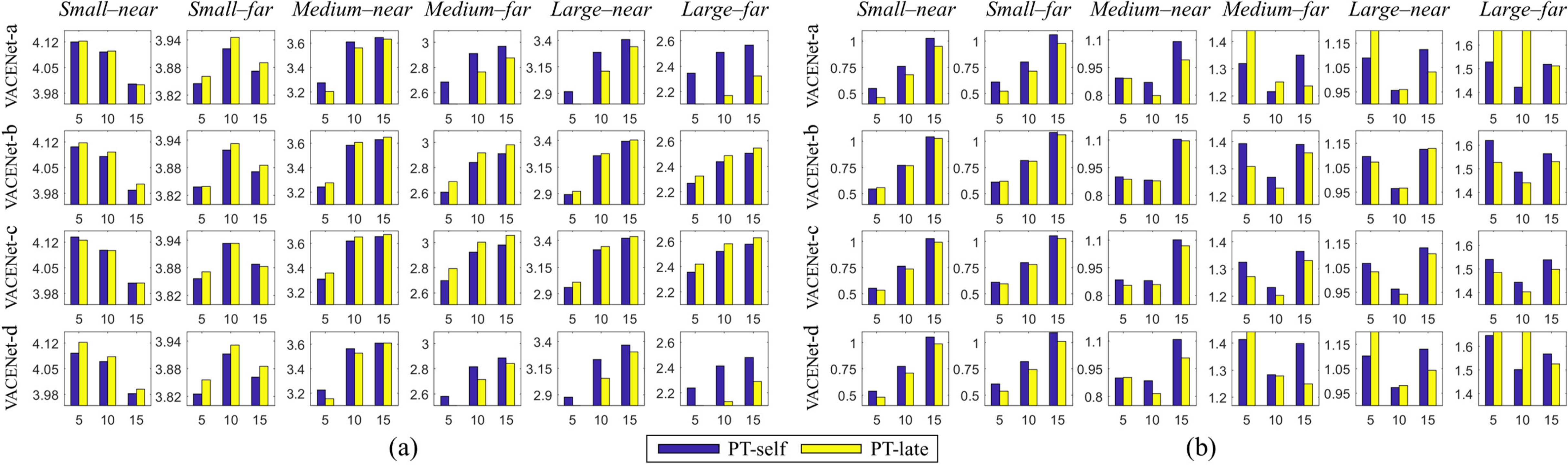}}
		\caption{
			Performance comparison of the simplified VACE-WPE systems built with different pre-training strategies (i.\,e., PT-self and PT-late) and VACENet structures (i.\,e., VACENet-\{a,\,b,\,c,\,d\}):
			(a) PESQ and (b) CD.
			The horizontal axis denotes the LP filter order, $K$.
			$K_\textrm{trn}$ was set to 10 during fine-tuning.
		}
		\label{fig:plot_C}
	\end{center}
\end{figure*}

\subsection{VACENet Architecture and Pre-training Methods} \label{sec5:C}
As briefly mentioned in Section \ref{sec5:A:2}, we observed a resemblance between the virtual signal and late reverberation to an extent.
Fig.\,\ref{fig:plot_AB_sf_late} shows the spectrograms and waveforms of the virtual signals and those of the oracle late reverberation component of the observed signal; 
the first two were generated using the VACE-WPE \cite{vace_wpe:is20} and its simplified version, respectively.
As seen in the figure, all these signals are clearly different from the reverberant input signals ($X_1$ and $X_2$) depicted in Fig.\,\ref{fig:plot_AB_sf_sigfilt}, yet are partially similar to each other; 
for example, the waveforms in the time-domain or the temporal distribution of ``hot" regions of the spectrograms.
Inspired by this, we proposed to pre-train the VACENet to estimate the late reverberation component of the observed signal, as described in Section \ref{sec3:C:4}.

Fig.\,\ref{fig:plot_C} compares the PESQ and CD measures obtained from the different VACE-WPE systems, each of which is distinguished by the pre-training strategy employed and the VACENet structure;
details of the four different VACENet models can be found in Fig.\,\ref{fig:vacenet_architectures} and Table \ref{tab:vacenet_params} in Section \ref{sec3:C:2}.
In the figure, the results for $K\in\{20,\,25\}$ were omitted because the simplified VACE-WPE revealed unfavorably high CD values with nearly consistent PESQ and SRMR scores (see Fig.\,\ref{fig:plot_B}). 
First, focusing on the impact of the new pre-training strategy on the four VACENet models, the VACE-WPE systems built with the VACENet-\{b,\,c\} models revealed noticeable improvement via adoption of the PT-late method in both the medium and large room conditions; 
they exhibited negligible difference in the small room conditions.
Moreover, between the VACENet-b and VACENet-c, the latter was overall superior to the former.
In contrast, when the PT-late strategy was introduced to the systems built with the VACENet-\{a,\,d\}, the performance was marginally improved in the small rooms, but was substantially degraded in the \textit{Medium-far}, \textit{Large-near}, and \textit{Large-far} conditions, with regard to either the PESQ or CD measure.
This may be possibly due to their distinctive structure, where they employ either a shared or separate stream for both of the encoder and decoder, as depicted in Fig.\,\ref{fig:vacenet_architectures}.

Next, comparing the VACENet structures initialized with the PT-self method, the VACENet-a and VACENet-c, both of which have a shared-stream decoder for modeling the RI components of the virtual signal, broadly outperformed the others in terms of both the PESQ and CD metrics.
Meanwhile, VACENet-d exhibited the worst performance in the \textit{Medium-far} and the large room conditions, under both the PT-self and PT-late strategies.

To summarize, among the eight different VACE-WPE systems under evaluation, the combination of the VACENet-c structure and the PT-late strategy for initialization showed the best performance.

\begin{figure*}[!t]
	\begin{center}
		\centerline{\includegraphics[width=5.3in]{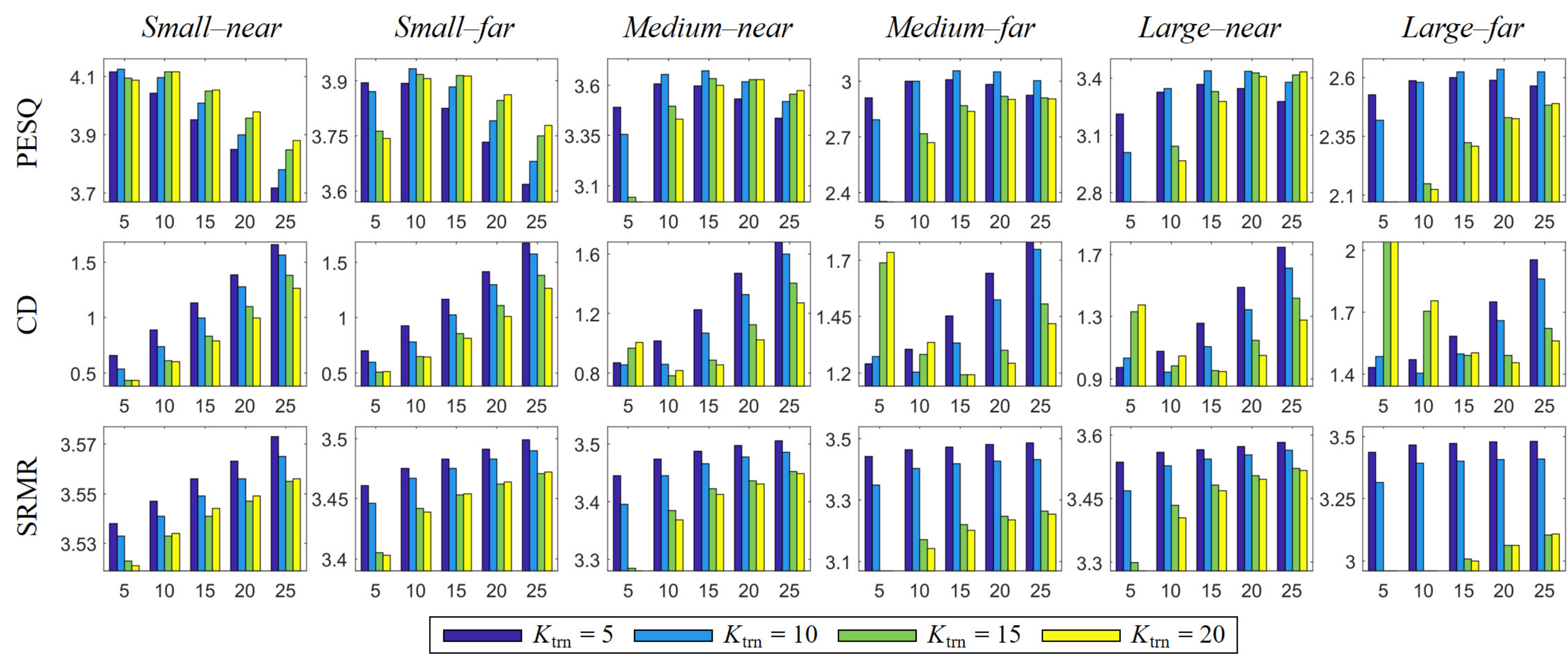}}
		\caption{
			Performance comparison of the simplified VACE-WPE systems fine-tuned with different LP orders, $K_\textrm{trn}\in\{5,\,10,\,15,\,20\}$.
			The VACENet-c model pre-trained using the PT-late method was adopted for fine-tuning.
			The horizontal axis denotes the LP filter order, $K$.
		}
		\label{fig:plot_D}
	\end{center}
\end{figure*}

\begin{figure}[t]
	\begin{center}
		\centerline{\includegraphics[width=\linewidth]{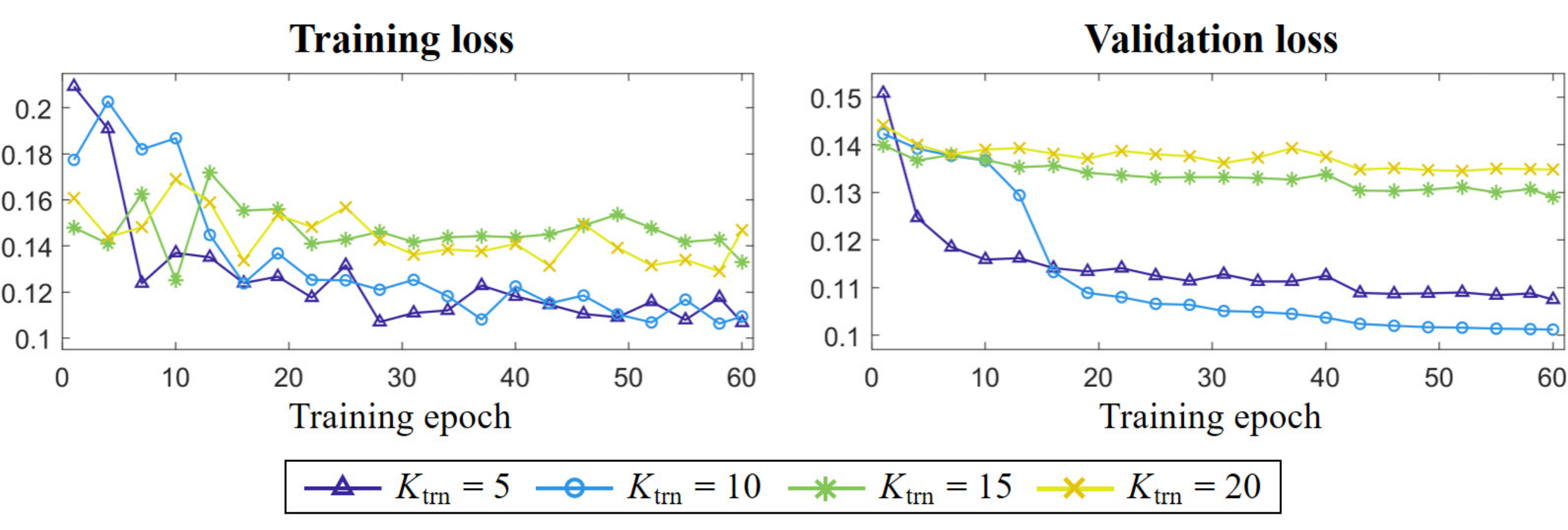}}
		\caption{
			Training and validation losses observed during fine-tuning the simplified VACE-WPE with the different LP orders, $K_\textrm{trn}\in\{5,\,10,\,15,\,20\}$.
			The validation loss was calculated with $K=15$, and the loss values were depicted for every third epoch for visual clarity.
		}
		\label{fig:plot_loss}
	\end{center}
\end{figure}

\subsection{Effect of the LP Order Set During the Fine-tuning} \label{sec5:D}
In this subsection, we investigate the effect of different LP orders set during the fine-tuning of the VACENet.
Based on the analysis in Section \ref{sec5:C}, we constructed a simplified VACE-WPE on top of the VACENet-c model initialized using the PT-late method.
Fig.\,\ref{fig:plot_D} shows the performance of the VACE-WPE systems fine-tuned with the different values of the LP orders, $K_\textrm{trn}\in\{5,\,10,\,15,\,20\}$, in terms of the PESQ, CD, and SRMR metrics.
Notably, the systems trained with relatively large LP orders of $K_\textrm{trn}\in\{15,\,20\}$ tend to severely fail in the medium and large room conditions, when evaluated using the smaller LP orders of $K\in\{5,\,10\}$.
In contrast, under the same test conditions, the systems built with relatively small LP orders of $K_\textrm{trn}\in\{5,\,10\}$ showed favorable trade-offs between the PESQ and CD metrics measured with $K=5$ and those measured with $K=10$, while exhibiting adversely high CD values for $K\in\{15,\,20,\,25\}$.
These two contrasting trends mildly indicate that the VACENet, pre-trained using the PT-late strategy, is in fact fit to generate the virtual signal that is basically the most effective as the auxiliary input when the back-end WPE algorithm operates with the LP order close to that employed in the fine-tuning stage.
This may be attributed to training the VACENet in an end-to-end manner within the WPE dereverberation framework, where the algorithm is restricted to operate with a fixed LP order.
However, the VACE-WPE systems trained with $K_\textrm{trn}\in\{15,\,20\}$, even when evaluated using the matched LP orders of $K\in\{15,\,20\}$, failed to achieve high PESQ and SRMR scores in the \textit{Medium-far} and \textit{Large-far} conditions.
This is explained in Fig.\,\ref{fig:plot_loss}, which visualizes 
the training and validation losses observed during the fine-tuning of the four different VACE-WPE systems;
the validation loss was computed on a small, separate validation set using $K=15$.
It can be seen from the figure that, unlike the systems trained with $K_\textrm{trn}\in\{5,\,10\}$, those trained with $K_\textrm{trn}\in\{15,\,20\}$ fail to sufficiently reduce both the training and validation losses.
Furthermore, comparing the two systems trained with $K_\textrm{trn}=5$ and $K_\textrm{trn}=10$, the former certainly experienced a faster convergence than the latter.
These observations indicate that generating virtual input signals from scratch against the dual-channel WPE operating with relatively large LP orders is difficult, possibly because the degrees of freedom of the relevant matrices presented in Eqs. (\ref{eqn:wpe_step2})\,--\,(\ref{eqn:wpe_step2-3}) increases with the LP order.
Nonetheless, it is quite impressive that the VACE-WPE fine-tuned with $K_\textrm{trn}=5$ performed well in the large room conditions, even when evaluated using relatively small LP orders of $K\in\{5,\,10\}$.

Meanwhile, in the small room conditions, the systems trained with $K_\textrm{trn}\in\{15,\,20\}$ were comparable or marginally superior to those trained with $K_\textrm{trn}\in\{5,\,10\}$ in terms of the PESQ and CD measures, with slightly lower SRMR scores.

\subsection{Results in Noisy Reverberant Conditions} \label{sec5:E}
%
In this subsection, the performance of the VACE-WPE is verified under noisy reverberant test conditions.
Both the LPSNet and VACENet-c models were trained using the \textit{TrainSimuNoisy} dataset as described in Sections \ref{sec4:D} and \ref{sec4:E}.
The PT-late strategy was adopted to pre-train the VACENet.
Herein, the early arriving speech plus noise was employed as the target signal for training the LPSNet and VACENet, as the WPE algorithm is only capable of blind dereverberation, but not explicitly designed for noise removal.
Based on the observation from Fig.\,\ref{fig:plot_loss}, we fine-tuned the VACENet by gradually increasing the LP filter order, $K_\textrm{trn}$, as the training progresses.
More specifically, for every single mini-batch, $K_\textrm{trn}$ was randomly chosen within the set $S_{K} = \{ K \,|\, K_\textrm{trn}^\textrm{lower} \leq K \leq K_\textrm{trn}^\textrm{upper} \} \subset \mathbb{Z}^\mathbf{+}$, 
and the optimization was performed using the selected LP order;
$K_\textrm{trn}^\textrm{lower}$ was fixed at 4, and $K_\textrm{trn}^\textrm{upper}$ was initially set to 6 and increased to 9, 12, 15, 18, and 21 after the 15th, 25th, 35th, 44th, and 52nd epochs, respectively.

\begin{figure}[t]
	\begin{center}
		\centerline{\includegraphics[width=3.2in]{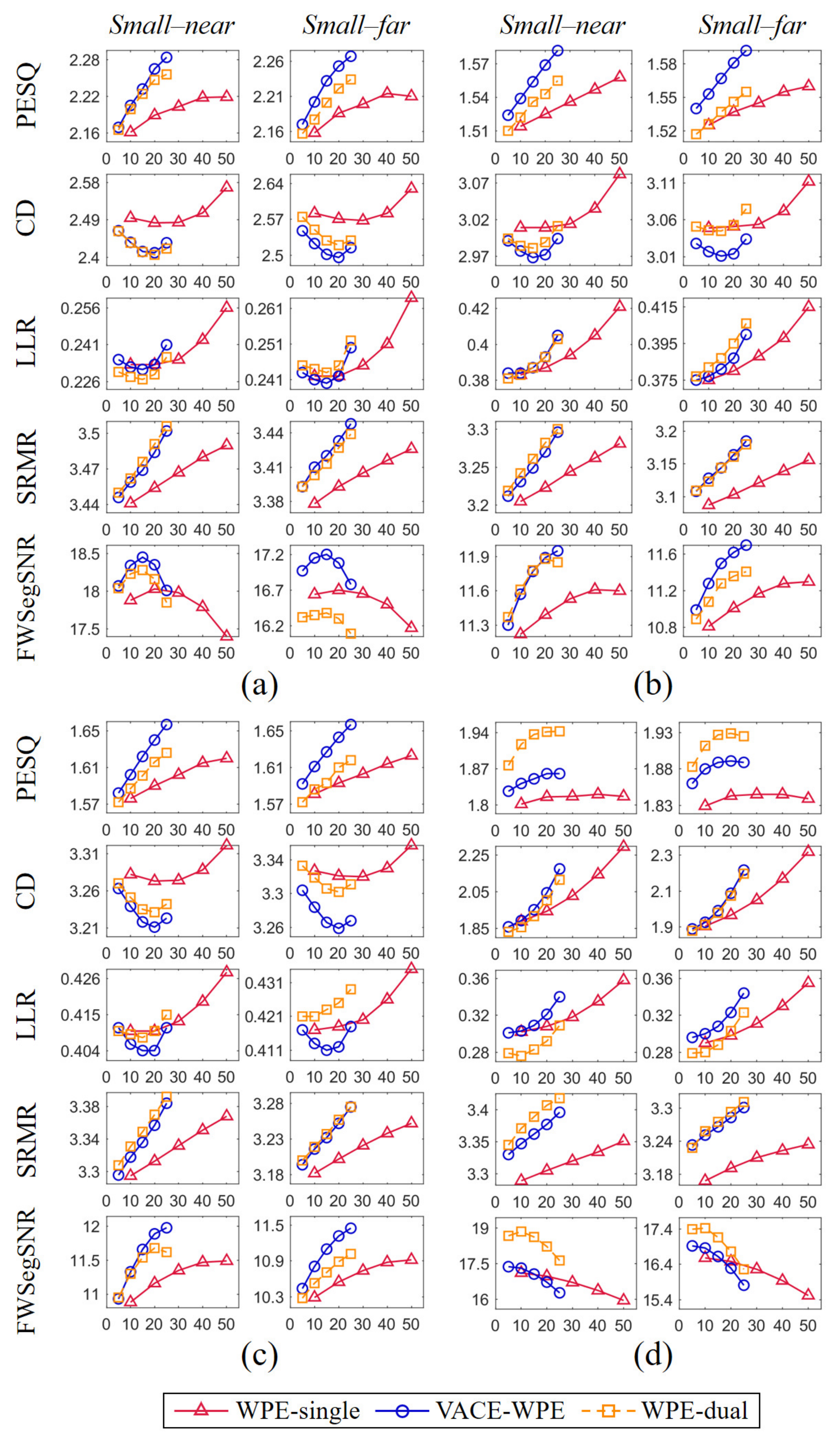}}
		\caption{
			Speech dereverberation performance on \textit{TestRealNoisy} in the small room environment: 
			(a) air conditioner, 
			(b) babble, 
			(c) factory, and
			(d) music.
			The horizontal axis denotes the LP filter order, $K$.
		}
		\label{fig:plot_E_small}
	\end{center}
\end{figure}

\begin{figure*}[!t]
	\begin{center}
		\centerline{\includegraphics[width=6.7in]{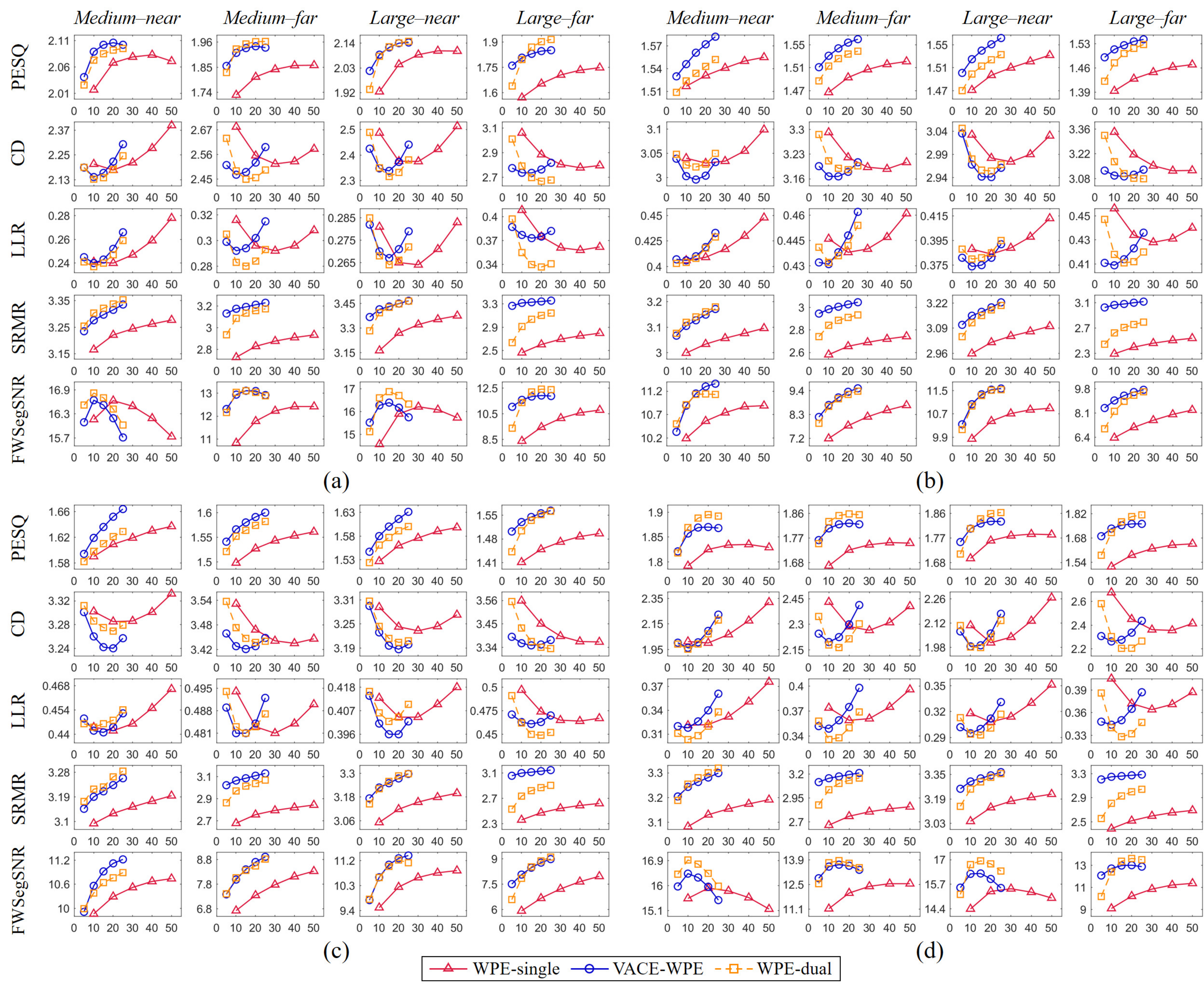}}
		\caption{
			Speech dereverberation performance on \textit{TestRealNoisy} in the medium and large room environments: 
			(a) air conditioner, 
			(b) babble, 
			(c) factory, and
			(d) music.
			The horizontal axis denotes the LP filter order, $K$.
		}
		\label{fig:plot_E_medium_large}
	\end{center}
\end{figure*}

The evaluation results on the \textit{TestRealNoisy} dataset are shown in Figs.\,\ref{fig:plot_E_small} and \ref{fig:plot_E_medium_large}, where the former demonstrates those measured in the small room environment and the latter in the medium and large rooms.
Comparing the single-channel WPE and VACE-WPE, it can be confirmed that the latter tends to exhibit operating points generally superior to those of the former in terms of all the evaluation metrics considered.
Similar to the results obtained in Section \ref{sec5:A}, the performance gap between the two algorithms further increased in the far-field speaking conditions, particularly with regard to the PESQ, SRMR, and FWSegSNR metrics.
Moreover, the VACE-WPE was also favorably comparable to the dual-channel WPE, revealing marginally better PESQ measures in the babble and factory noise conditions in various room environments and moderately higher SRMR scores in the \textit{Medium-far} and \textit{Large-far} conditions.
Interestingly, these SRMR scores measured with the different values of the LP order imply that the VACE-WPE is better capable of producing ``dry" signals than the dual-channel WPE using relatively small LP orders.
Finally, considering that there exists a mismatch between the clean speech corpus of \textit{TrainSimuNoisy} and that of \textit{TestRealNoisy}, it can be stated that the training of the VACE-WPE can generalize well to a larger corpus, instead of simply being overfit to a small-scale dataset.

\subsection{Speech Recognition Results on Real Recordings} \label{sec5:F}
In this subsection, we verify the performance of the various speech dereverberation methods as the front-end for the automatic speech recognition (ASR) task.
Specifically, we followed the protocol for the ASR task of the VOiCES Challenge 2019 \cite{voices_channelge_2019_eval_plan,voices_challenge_2019}, a recent benchmark on far-field ASR in challenging noisy reverberant room environments.
The challenge provides two different sets of utterances for the system development and evaluation, namely the ``dev" and ``eval" sets \cite{voices_channelge_2019_eval_plan,voices_challenge_2019};
each set consists of a small portion of the VOiCES corpus \cite{voices_corpus_2018}.
The VOiCES corpus is a re-recorded subset of the LibriSpeech dataset \cite{librispeech}, and the re-recording was performed using twelve microphones of different types and locations in the presence of background noise, for example, fan, babble, music, and television \cite{voices_corpus_2018}.
To build the baseline ASR system, we used an open source script\footnote{{https://github.com/freewym/kaldi-voices}} that partially implements the system described in \cite{voices19_jhu_asr} based on the Kaldi \cite{kaldi_toolkit} toolkit.
The acoustic model\footnote{kaldi/egs/librispeech/s5/local/chain/tuning/run\_cnn\_tdnn\_1a.sh} 
was built using the modified LibriSpeech-80h dataset \cite{voices_channelge_2019_eval_plan,voices_challenge_2019} after applying the standard data augmentation and speed perturbation \cite{kaldi_rirs} provided by the Kaldi recipes \cite{kaldi_toolkit}; 
40-dimensional log-mel-filterbank energies, extracted with a 25 ms window and 10 ms hop sizes, were used as the input acoustic features.
A 3-gram statistical language model constructed using the transcripts of the training utterances was employed for decoding.

Tables \ref{tab:voices_srmr} and \ref{tab:voices_wer} present the SRMR scores and word error rate (WER) obtained using the different speech dereverberation methods, respectively.
For the single-channel WPE and VACE-WPE, the LP filter order, $K$, was set to 80 and 35, respectively;
further increasing $K$ did not improve the performance of both algorithms significantly.
As shown in the tables, besides the single-channel WPE, two different fully neural speech dereverberation models, namely the LPSNet-Drv and VACENet-c-Drv, were also under comparison.
More specifically, the LPSNet-Drv was implemented by simply combining the dereverberated magnitude spectra, estimated from the trained LPSNet, with the phase spectra of the reverberant observation.
The VACENet-c-Drv was obtained by training a neural network, whose structure is identical to the VACENet-c, to estimate the RI components of the early arriving speech plus noise.
These models allow to make a direct comparison between \romannumeral 1) employing the neural network for directly estimating the early arriving speech component and \romannumeral 2) employing the neural network for the virtual signal generation instead and subsequently let the pre-trained dual-channel neural WPE perform the dereverberation.
Table \ref{tab:voices_srmr} illustrates that the VACE-WPE and VACENet-c-Drv reveal significantly higher SRMR scores relative to the other methods and are comparable with each other.
However, as shown in Table \ref{tab:voices_wer}, the single-channel WPE achieved the lowest WER in both sets, followed by the VACE-WPE that revealed slightly worse performance;
both the LPSNet-Drv and VACENet-c-Drv failed to reduce the WER.
Accordingly, it can be stated that the proposed VACE-WPE can achieve a great balance between the objective speech quality improvement and front-end processing for the ASR task in terms of dereverberation.

Table \ref{tab:voices_lattice_interp} further presents the results obtained after performing lattice interpolation \cite{lattice_interp:icassp12} on top of the ASR output lattices generated using the single-channel WPE front-end and those using the VACE-WPE;
the scaling factor, $\lambda$, was varied from 0.1 to 0.9.
Absolute decrements of 0.3\% and 0.9\% in WER, achieved on the ``dev" and ``eval" sets, respectively, indicate that the single-channel WPE and VACE-WPE can be complementary as the speech dereverberation front-end for the ASR task.

\begin{table}[]
	\caption{SRMR scores measured on the VOiCES Challenge 2019 \cite{voices_challenge_2019} dataset}
	\centering
	\begin{tabular}{@{}cccccc@{}}
		\toprule
		Method & \begin{tabular}[c]{@{}c@{}}Raw\\ signal\end{tabular} & \begin{tabular}[c]{@{}c@{}}WPE-single\\ \,($K=80$)\end{tabular} & \begin{tabular}[c]{@{}c@{}}VACE-WPE\\ ($K=35$)\end{tabular} & \begin{tabular}[c]{@{}c@{}}LPSNet-\\ Drv\end{tabular} & \begin{tabular}[c]{@{}c@{}}VACENet-c-\\ Drv\end{tabular} \\ \midrule
		dev    & 2.30  &  2.80  & \textbf{3.10}  & 2.81  & \textbf{3.12}  \\
		eval   & 2.07  &  2.59  & \textbf{3.07}  & 2.67  & \textbf{3.06}  \\ \bottomrule
	\end{tabular}
	\label{tab:voices_srmr}
\end{table}

\begin{table}[]
	\caption{WER(\%) measured on the VOiCES Challenge 2019 \cite{voices_challenge_2019} dataset}
	\centering
	\begin{tabular}{@{}cccccc@{}}
		\toprule
		Method & \begin{tabular}[c]{@{}c@{}}Raw\\ signal\end{tabular} & \begin{tabular}[c]{@{}c@{}}WPE-single\\ \,($K=80$)\end{tabular} & \begin{tabular}[c]{@{}c@{}}VACE-WPE\\ ($K=35$)\end{tabular} & \begin{tabular}[c]{@{}c@{}}LPSNet-\\ Drv\end{tabular} & \begin{tabular}[c]{@{}c@{}}VACENet-c-\\ Drv\end{tabular} \\ \midrule
		dev    & 24.2  &  \textbf{21.3}  & \textbf{21.5}  & 25.7  & 26.1  \\
		eval   & 30.0  &  \textbf{24.9}  & \textbf{25.1}  & 30.4  & 31.0  \\ \bottomrule
	\end{tabular}
	\label{tab:voices_wer}
\end{table}

\begin{table}[]
	\caption{
		WER(\%) after performing lattice interpolation \cite{lattice_interp:icassp12} between the ASR output lattices generated using the single-channel WPE and those using the VACE-WPE.
		$\lambda$ was applied to the former and $1-\lambda$ to the latter
	}
	\centering
	\begin{tabular}{@{}x{0.493cm}x{0.493cm}x{0.493cm}x{0.493cm}x{0.493cm}x{0.493cm}x{0.493cm}x{0.493cm}x{0.493cm}x{0.493cm}@{}}
		\toprule
		$\lambda$ & 0.1  & 0.2  & 0.3  & 0.4  & 0.5  & 0.6  & 0.7  & 0.8  & 0.9  \\ \midrule
		dev       & 21.3 & 21.1 & 21.1 & \textbf{21.0} & \textbf{21.0} & \textbf{21.0} & \textbf{21.0} & 21.1 & 21.2 \\
		eval      & 24.4 & 24.3 & 24.1 & \textbf{24.0} & \textbf{24.0} & \textbf{24.0} & \textbf{24.0} & 24.2 & 24.3 \\ \bottomrule
	\end{tabular}
	\label{tab:voices_lattice_interp}
\end{table}

\section{Conclusions} \label{sec6}
In this study, we first investigated the properties of the VACE-WPE system via ablation studies, which led to the introduction of a simplified architecture and new strategies for training the neural network for the VACE.
Based on these findings, the performance of the VACE-WPE was further examined with regard to \romannumeral 1) objective quality of the dereverberated speech under noisy reverberant conditions and \romannumeral 2) ASR results measured on real noisy reverberant recordings.
Experimental results and analysis indicate that the neural-network-based virtual signal generation followed by the modified neural WPE back-end can provide an implementation of an effective speech dereverberation algorithm in a single-microphone offline processing scenario.
%
%


%





\ifCLASSOPTIONcaptionsoff
  \newpage
\fi



%

\bibliographystyle{IEEEtran}
\bibliography{vace_wpe_arxiv}

\end{document}